\documentclass[aps,reprint,pre,amsfonts,amsmath,superscriptaddress]{revtex4-2}

\usepackage{graphicx}
\usepackage{times}
\usepackage{hyperref}
\usepackage[caption=false]{subfig}

\begin{document}

\title{Hyperoptimized approximate contraction of tensor networks for rugged-energy-landscape spin glasses on periodic square and cubic lattices}

\author{Adil A. Gangat}
\affiliation{Physics \& Informatics Laboratories, NTT Research, Inc., Sunnyvale, CA 94085}
\affiliation{Division of Chemistry and Chemical Engineering, California Institute of Technology, Pasadena, CA 91125}
\author{Johnnie Gray}
\affiliation{Division of Chemistry and Chemical Engineering, California Institute of Technology, Pasadena, CA 91125}

\date{\today}
\begin{abstract}
Obtaining the low-energy configurations of spin glasses that have rugged energy landscapes is of direct relevance to combinatorial optimization and fundamental science.  Search-based heuristics have difficulty with this task due to the existence of many local minima that are far from optimal.  The work of [M. M. Rams \textit{et al.}, Phys. Rev. E \textbf{104}, 025308 (2021)] demonstrates an alternative that can bypass this issue for spin glasses with planar or quasi-planar geometry: sampling the Boltzmann distribution via approximate contractions of tensor networks.  The computational complexity of this approach is due only to the complexity of contracting the network, and is therefore independent of landscape ruggedness.  Here we initiate an investigation of how to take this approach beyond (quasi-)planar geometry by utilizing hyperoptimized approximate contraction of tensor networks [J. Gray and G. K.-L. Chan, Phys. Rev. X \textbf{14}, 011009 (2024)].   We perform tests on the periodic square- and cubic-lattice, planted-solution Ising spin glasses generated with tile planting [F. Hamze \textit{et al.}, Phys. Rev. E \textbf{97}, 043303 (2018)] for up to 2304 (square lattice) and 216 (cubic lattice) spins.  For a fixed bond dimension, the time complexity is quadratic.  With a bond dimension of only four, over the tested system sizes the average relative energy error in the most rugged instance class remains at $\sim$1\% (square lattice) or $\sim$10\% (cubic lattice) of optimal.  In less rugged instances the solution is always optimal for the square lattice and either optimal or within $\sim$1\% for the cubic lattice.  These results suggest that further development of optimization methods based on tensor-network representations of spin glass partition functions may be fruitful, especially given that such methods are not limited to the Ising (i.e., binary) or two-body (i.e., quadratic) settings.
\end{abstract}
\maketitle

\section{Introduction}
\label{sec:intro}

Discrete combinatorial optimization problems and glassy materials with quenched disorder are both often modeled with classical spin Hamiltonians such that the interaction energies between the spins are disordered.  The simplest such ``spin glass" model is the disordered classical Ising model:
\begin{equation}
H = \frac{1}{2}\sum_{i\neq j} J_{ij}s_is_j,
\label{eqn:Ham}
\end{equation}
where $i$ and $j$ are graph vertex labels; $J_{ij}$ is a random, real scalar for the interaction energy between spins at vertices $i$ and $j$; and $s_i=\pm1$.  Obtaining good solutions to discrete optimization problems and also understanding the low-temperature behavior of glassy systems then entails obtaining the low-energy (a.k.a. near-optimal) and lowest-energy (a.k.a. optimal) configurations of models like Eq. (\ref{eqn:Ham}) \footnote{The definition of ``near-optimal'' is problem-specific. For example, for Max-Cut problems it is in some cases desirable to have solutions within $\sim12\%$ of optimal and in other cases within $\sim6\%$ \cite{mohseni2022ising}.}.

Specialized hardware platforms for obtaining near-optimal and optimal configurations of spin glasses are under active development \cite{cai2020power, mohseni2022ising, nguyen2024entropy, zhang2024review, king2024computational}.  Although such machines display an average time complexity that is superpolynomial for obtaining optimal solutions \cite{mohseni2022ising} (but see Refs. \cite{traversa2017polynomial, bernaschi2024quantum, zhang2024cyclic, ghosh2024exponential} for the theoretical possibility of average polynomial time complexity), the hope is that the smaller constant factors in the scaling that are possible with such machines will one day yield superior performance compared to classical CPU-based methods for industrial optimization problems at scale.  A recent milestone of particular note toward this goal is the experimental confirmation by D-Wave, in the context of a 5,000-qubit simulation of a three-dimensional quantum spin glass, of a scaling advantage of quantum annealing over both simulated annealing and (Monte Carlo) simulated quantum annealing for the solution quality (i.e., spin configuration energy distance from the ground state) versus the annealing time \cite{king2023quantum}.  Regarding computing near-optimal solutions (a.k.a. approximate optimization), a recent work \cite{bauza2024scaling} showed that quantum annealing can obtain an average polynomial time complexity for solutions within 1\% error from optimal of a quasi-two-dimensional classical spin glass, and further that it can do so with a scaling exponent that is smaller than that of the leading classical heuristic (parallel tempering Monte Carlo with isoenergetic cluster moves \cite{zhu2015efficient}).

But for both exact and approximate optimization, there is much room left to improve such specialized hardware platforms.  For example, densely connected Ising Hamiltonians are the typical case in industrial applications \cite{konz2021embedding}, so the sparse connectivity between qubits in the current D-Wave hardware requires an embedding of the dense problem, which incurs a quadratic overhead in the number of spins \cite{konz2021embedding} that dramatically impacts its scaling performance \cite{hamerly2019experimental}.  As another example, the quantum approximate optimization experiments in Ref. \cite{harrigan2021quantum} show solution quality that remains constant with increasing problem size for spin glass problems with geometry that is native to their superconducting-qubit processor's planar geometry, but decreasing solution quality with increasing problem size for geometrically non-native problems that are embedded (a.k.a. compiled) into their processor's geometry.  Therefore, further development of classical algorithms for exact and approximate optimization remains relevant, as they can compute larger scale and more complex problems than current specialized machines, and also serve as a way of benchmarking the performance of the specialized machines as they are further developed.

As for classical digital algorithms, many are based on Markov Chain Monte Carlo (MCMC) \cite{krauth2006statistical}, which performs an annealing-based stochastic search over the low-temperature free-energy landscape in configuration space for the global minima and also local minima that are close in energy to the global minima.  However, in various spin-glass cases the ruggedness of the landscape is such that these methods become exponentially slower with increasing system size \cite{machta2009strengths,ciarella2023machine,montanari2006rigorous}.  There are some proposals to accelerate MCMC methods with assistance from machine learning, but Ref. \cite{ciarella2023machine} shows that these proposals fail in the presence of rugged energy landscapes.  In fact, Ref. \cite{ciarella2023machine} concludes that rugged-energy-landscape spin glass problems may be intrinsically difficult for all machine learning algorithms, and we are not aware of any counterexamples thus far.  In fact, the difficulty that MCMC-based methods face in dealing with rugged energy landscapes in the configuration space of spin glass phases has been suggested as indicative of a place to look for quantum advantage \cite{jauma2024exploring}.

On the other hand, there are classical digital algorithms that are based on simulation of dynamical systems, such as the simulated coherent Ising machine \cite{tiunov2019annealing,leleu2021scaling} and simulated bifurcation machine \cite{goto2019combinatorial,goto2021high}, but these also amount to a search over the free-energy landscape in configuration space for optimal and near-optimal minima, and can therefore also suffer a steep performance drop (in terms of time to solution) with increasing landscape ruggedness, as observed for example in the supplementary materials of Ref. \cite{leleu2021scaling}.

For industrial applications it is often necessary to obtain near-optimal solutions to combinatorial optimization problems at very large system sizes \footnote{It is often the case in industrial contexts that the optimal solution is not necessary; \textit{reduced cost} solutions are often sufficient.}, making polynomial time complexity highly desirable.  It is important then to find numerical methods that can approximately solve rugged-energy-landscape classical spin glasses in polynomial time.  Tensor networks \cite{orus2014practical,banuls2023tensor} offer a route to approximately sampling, in polynomial time, the low-temperature Boltzmann distribution of classical spin glasses without traversing the free energy landscape in configuration space, and may thus offer a  way to bypass the problems presented by rugged energy landscapes without the use of quantum hardware.  This route is now possible due to the following key developments:  First, the pioneering work of Nishino \cite{nishino1995density, nishino1995product, nishino1998density} showed how to use tensor-network algorithms for \textit{homogenous} two- and three-dimensional classical lattice models.  Then,  Murg \textit{et al.} presented a way in Ref. \cite{murg2005efficient} to use tensor networks to deterministically perform efficient evaluation of partition functions of finite, \textit{inhomogeneous} two-dimensional classical lattice models with open boundaries (although they did not apply their approach to a spin glass).  This latter approach is also reviewed in Ref. \cite{verstraete2008matrix}.  Their idea entails a lattice of low-rank tensors that reflects the geometry of the physical lattice; an approximation of the partition function is obtained in polynomial time by approximately contracting all of the tensors in a row-to-row (or column-to-column) fashion by treating the rows (or columns) as matrix product states (MPSs) or matrix product operators (MPOs), which are one-dimensional tensor networks.  The approximation arises due to setting an upper bound on the size of the indices of the intermediate tensors that arise during the contraction.  Since the computational complexity of this approximate method is independent of the elements in the tensors, it is independent of the ruggedness of the energy landscape. The subsequent work of Ref. \cite{rams2021approximate} develops the idea of Murg \textit{et al.} further by demonstrating how to sample actual spin configurations from the low-temperature Boltzmann distribution of spin glasses via such approximate contractions.

We note some limitations of the work in Ref. \cite{rams2021approximate}: It is only able to natively (i.e., without embedding) handle spin glasses on graphs that are either planar or quasi-planar (e.g., the chimera graph), which prevents a native treatment of spin glass problems with non-planar geometry. Though non-planar problems can be embedded into quasi-planar graphs, this entails a polynomial overhead in the number of spins \cite{konz2021embedding}.  Further, while Ref. \cite{rams2021approximate} demonstrates good solution quality with small-index-size tensors for large, rugged-landscape spin glasses that are native to the chimera graph, they do not test against problems with non-planar geometry that have been embedded into the chimera geometry.  This is significant because tests carried out in Ref. \cite{jalowiecki2021brute} show that non-natively representing the Boltzmann distribution of a complete-graph model by using an MPS requires the size of the tensors of the MPS to grow rapidly with system size if good solution quality is desired.  Also, the above-mentioned experiment \cite{harrigan2021quantum} that found decreasing solution quality with increasing problem size for geometrically non-native problems that are embedded into the native planar geometry of a superconducting-qubit processor makes it doubtful that the quasi-planar tensor-network method in Ref. \cite{rams2021approximate} would yield good solutions in polynomial time for problems that are not \textit{natively} planar or quasi-planar.

Ref. \cite{liu2021tropical} presents a different tensor network approach for sampling from the partition function, based on tropical algebra, that can natively handle non-planar geometries and that guarantees the optimal solution with a time to solution that is independent of landscape ruggedness.  However, the time complexity of this exact approach is (unsurprisingly) superpolynomial in the system size, and due to the use of tropical algebra there is as yet no established way of incorporating approximations into that approach to reduce the time complexity to polynomial.  The recent work in Ref. \cite{lanthier2024tensor} presents a still different tensor network method that is tailored to the exact partition function computation of the \textit{p}-spin model (which can exhibit rugged energy landscapes) and can also natively handle non-planar geometry, yet again in some regimes has superpolynomial time complexity.

Thus it is important to find ways of extending the polynomial-time approach of Ref. \cite{rams2021approximate} to be able to natively handle non-planar geometries.  Strategies for approximately contracting tensor networks with arbitrary geometry do exist \cite{pan2020contracting, gray2024hyperoptimized, ma2024approximate}, and thereby provide a potential route for achieving this. Here we initiate the exploration of this route by testing the ability of the approximate tensor-network contraction method in Ref. \cite{gray2024hyperoptimized} to sample the low-temperature Boltzmann distributions of spin glasses on the following non-planar geometries: square and cubic lattices with periodic boundary conditions \footnote{Ref. \cite{gray2024hyperoptimized} shows that even in cases where an MPS-based contraction, such as the one used in Ref. \cite{rams2021approximate}, may be applied, the method of Ref. \cite{gray2024hyperoptimized} can significantly outperform MPS-based contraction in terms of both memory and time costs.}.  We use planted-solution models whose instances span distinct classes of energy landscape ruggedness.  For the square lattice we find the solution is always optimal except for in the most rugged class of instances, where we find an energy error of at most 3\%.  For the three ruggedness classes of the cubic lattice, we find an energy error of at most 12\% in the most rugged class, and 0\% and about 3\% in the other two.  These results are all obtained with the same polynomial time complexity \footnote{The (assumed) distinction between P and NP does not preclude the existence of \textit{heuristic} polynomial-time numerical methods for rugged-energy-landscape spin glasses---methods that have no performance guarantee regarding solution quality but in practice yield good solutions at least in some instances \cite{mohseni2022ising}}.  

\section{Method}
\label{sec:method}
We follow the general idea in Ref. \cite{rams2021approximate} for representing the Boltzmann distribution $p(\textbf{s})\sim\textrm{exp}[-\beta H(\textbf{s})]$, where $\textbf{s}$ is a spin configuration vector, as a contraction of a network of low-rank tensors; this gives access to any (unconditional or conditional) marginal probability of the system via trivial modifications of the tensor network.  Details of our tensor network formulation are given in Appendix \ref{sec:TN_construction}.  At a given value of $\beta$, our method computes the marginal probability distribution for a single spin by contracting a modified version of the original tensor network, updates the original tensor network by projecting that spin onto its most probable state, then computes the (conditional) marginal probability distribution of the next spin with the projected version of the original tensor network, and serially iterates this procedure for all remaining spins.  In this way the Boltzmann distribution for the entire spin glass at the chosen value of $\beta$ is sampled (though the sampling is not exact, due to the compression described below).  We repeat this at a series of values of $\beta$ and choose the lowest energy configuration as the best solution (limited numerical precision precludes the possibility of using arbitrarily large $\beta$).

Ref. \cite{rams2021approximate} performs a more sophisticated sampling via a branch and bound method, but we opt for the simpler strategy above since our focus here is only on providing evidence that hyperoptimized approximate tensor network contraction \cite{gray2024hyperoptimized} can, with a fixed polynomial time complexity, readily obtain solutions that are within about 10\% energy error from optimal for at least some \textit{non-planar} spin glass models with rugged energy landscapes.

During the contraction of any tensor network, the size of the indices of the intermediate tensors can grow exponentially.  The method in Ref. \cite{gray2024hyperoptimized}, which we employ in this work, can automatically limit, via compression, the index size to a maximum \textit{bond dimension} ($\chi$). For short-range spin models on lattices, using a constant value of $\chi$ limits the memory complexity for the tensor contractions to be constant in the number of spins ($N$).  The leading order memory complexity of the entire algorithm is therefore only that of the initial tensor network itself, which is $\mathcal{O}(N)$ for short-range models (see Appendix \ref{sec:TN_construction} for an example). For a fixed $\chi$, the time complexity in terms of $N$ is set by the number of tensor contractions. For short-range spin models on lattices, there are $\mathcal{O}(N)$ tensor contractions for each single-spin marginal.  Computing a spin configuration for a given $\beta$ with the method described above for all $N$ spins therefore has a time complexity $\mathcal{O}(N^2)$.  We note that the computations at different values of $\beta$ and $\chi$ are independent, and can therefore be done in parallel.

As explained in Ref. \cite{rams2021approximate}, while increasing $\beta$ theoretically reduces thermal fluctuations and therefore the $\chi$ that is needed to achieve a given solution quality (larger $\chi$ yields less compression and therefore more accuracy in principle), limited numerical precision means that in practice there is a finite value of $\beta$ that is optimal at any given value of $\chi$ (to see this, consider that the numerical range of the Boltzmann weights in Eq. (\ref{eqn:matrix}) in Appendix \ref{sec:TN_construction} monotonically increases with increasing $\beta$).  Finite numerical precision also means that increasing $\chi$ does not monotonically decrease the error at a fixed $\beta$.  Thus the values of $\chi$ and $\beta$ that optimize the solution quality are not known \textit{a priori}.  But as mentioned above, our purpose here is only to provide evidence that the general scheme is sufficiently performant in the context of rugged, non-planar spin glasses to be of interest for combinatorial optimization.  Here we say that this means that solutions within about $10\%$ of optimal should appear even at small values of fixed $\chi$, since computational cost monotonically increases with increasing $\chi$.  Hence we usually set $\chi=4$ and test a few values of $\beta$ to show that multiple solutions with energy error $\lesssim10\%$ do appear.

While the actual time cost of any tensor network contraction can depend strongly on the \textit{order} of the tensor contractions, the contraction algorithm in Ref. \cite{gray2024hyperoptimized} that we employ automatically chooses an optimal or near-optimal contraction order.  This search for the contraction order is done only once for a given lattice geometry, and not for every spin glass instance, so we do not include its cost in the computational complexity of the algorithm.  We implement the algorithm with the Python libraries \texttt{quimb} \cite{gray2018quimb} and \texttt{cotengra} \cite{gray2021hyper}.

\section{Benchmark Models}
Our testing uses the planted-solution Ising-model instances generated by the Chook library \cite{perera2020chook} with the tile planting method \cite{hamze2018near} on the square lattice and cubic lattice, both with periodic boundaries in all directions; in both cases the Hamiltonian takes the form in Eq. (\ref{eqn:Ham}).  Since the solutions are planted, the groundstate energy is exactly known.

\subsection{Square lattice tile planting}
\label{sec:TP2D_model}
A characterization of the landscape ruggedness of this model is contained in Ref. \cite{perera2020computational}.  The model contains four base classes of instances, designated as $C_1, C_2, C_3,$ and $C_4$.  The order of increasing (average) energy-landscape ruggedness of these is: $C_4, C_3, C_1,$ and $C_2$.  The results in Ref. \cite{perera2020computational} also indicate that this is the order of increasing average computational hardness for at least the following stochastic-search methods: population annealing, simulated annealing, and simulated quantum annealing.

The ground state is doubly degenerate; by default they are the ferromagnetic state (all $+1$ or all $-1$), but the Chook library has the option to scramble them with gauge transformations, which we enable for all problem instances.

\subsection{Cubic lattice tile planting}
A (partial) characterization of the landscape ruggedness of this model is contained in Ref. \cite{hamze2018near}.  The model contains three base classes of instances, designated as $F_2, F_4,$ and $F_6$, where $F_2$ is subdivided into $F_{21}$ and $F_{22}$, and $F_4$ into $F_{41}$ and $F_{42}$  The Chook library generates only problems in $F_{22}$, $F_{42}$, and $F_6$, and mixtures thereof.  The data in Ref. \cite{hamze2018near} shows that $F_6$ is computationally harder (i.e., more rugged) than both $F_{22}$ and $F_{42}$ for the stochastic-search methods of simulated annealing, parallel tempering, and population annealing.  The hardest instances of these problems are much harder for these stochastic search methods than problems with either bimodal or Gaussian disorder.

$F_{22}$, $F_{42}$, and $F_6$ all have a ferromagnetic groundstate, but $F_{22}$ and $F_6$ also have other groundstates: $F_{22}$ has a total of four groundstates (up to $\mathbb{Z}_2$ symmetry) and $F_6$ has a total of eight (up to $\mathbb{Z}_2$ symmetry).  For all problem instances, we enable the option in the Chook library to scramble the groundstates with gauge transformations.

\section{Results}
The simulations are done one instance at a time on an Apple M2 Ultra processor with 24 cores (16 performance cores and efficiency cores) and 128 GB of RAM.

\subsection{Square lattice tile planting}
\begin{figure}
    \centering
    \includegraphics[width=\columnwidth]{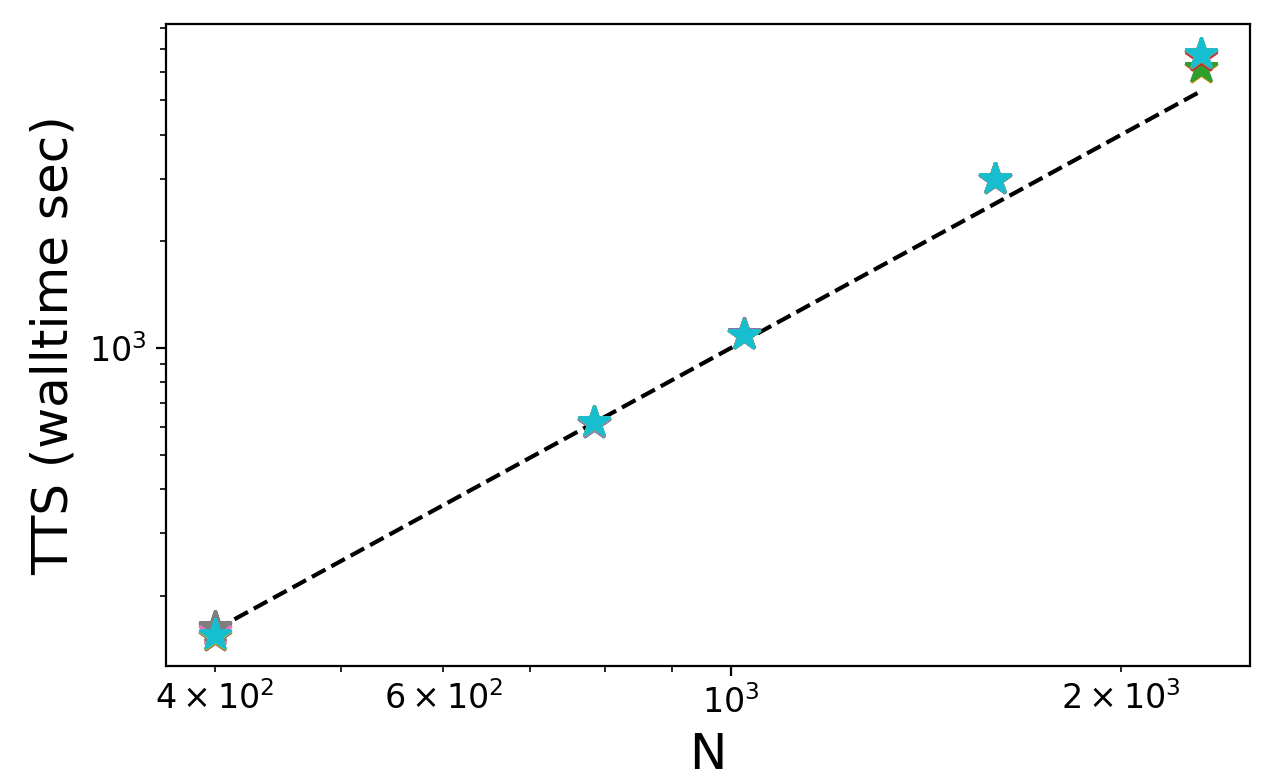}
    \caption{(color online). Square-lattice tile planting; wall-clock time to solution (stars) vs. total spins ($N=L^2$). Ten instances from $C_2$ at $L=20, 28, 32, 40$ and $48$ with $\chi=4$.  Dashed line indicates pure quadratic scaling.  The data is consistent with the predicted quadratic scaling for the leading order of the time cost.}
    \label{fig:TTS_C2}
\end{figure}

We test our method against the four base classes of the square-lattice model individually.  Within each base class, we test ten problem instances at sizes $L$ ($L^2$) = $20$ ($400$), $32$ ($1024$), and $48$ ($2304$).  The full data for solution quality vs. $\beta$ is presented in Appendix \ref{sec:TP2D_data}.  In Fig. \ref{fig:error_C2_48x48} we show that, with $\chi=4$, the method produces solutions that are within $2\%$ energy error for the most rugged base class ($C_2$) at $L=48$.  In Fig. \ref{fig:TTS_C2} we show the time to solution (TTS) vs. $N=L^2$ for $C_2$ with $\chi=4$, revealing the predicted quadratic scaling at constant $\chi$.

For a given problem instance, we consider the best solution as the one with the smallest energy error from the values of $\beta$ that we test.  We find the smallest error to always be zero in base classes $C_1, C_3, $ and $C_4$.  For $C_2$, in Fig. \ref{fig:minerror} we show that the average of such best solutions does not substantially change with increasing system size at $\chi=4$. 

\begin{figure}
    \centering
    \includegraphics[width=\columnwidth]{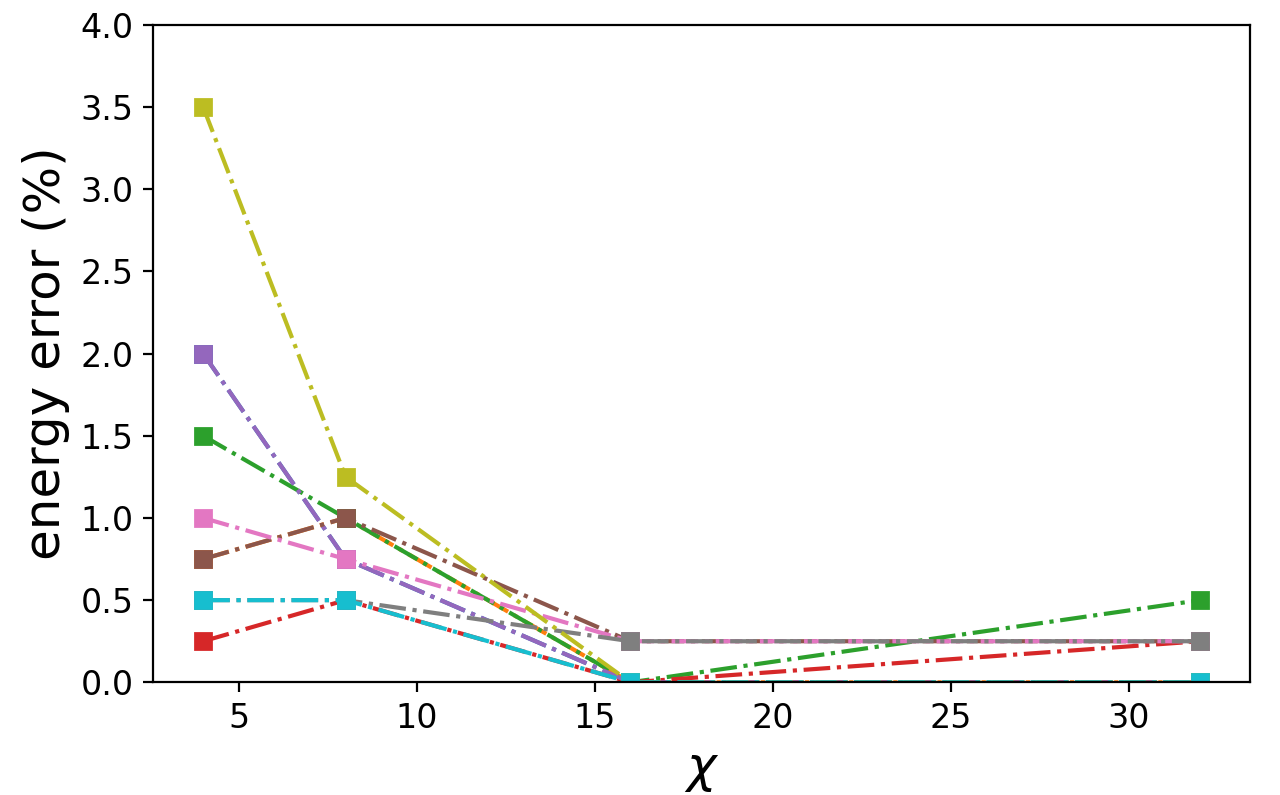}
    \caption{(color online). Square-lattice tile planting; error vs. $\chi$ at $L=20$ and $\beta=1$ for ten instances of the most rugged base class ($C_2$).  Data points belonging to the same instance are connected by dash-dotted lines.  The error can be diminished from that at $\chi=4$ by slightly increasing $\chi$.}
    \label{fig:error_vs_chi_C2_L20}
\end{figure}

In Fig. \ref{fig:error_vs_chi_C2_L20} we plot the energy error for all ten instances of $C_2$ at $L=20$ and $\beta=1$ as a function of $\chi$ to show that the error can be further decreased by only slightly increasing $\chi$.

\begin{figure}
    \centering
    \includegraphics[width=\columnwidth]{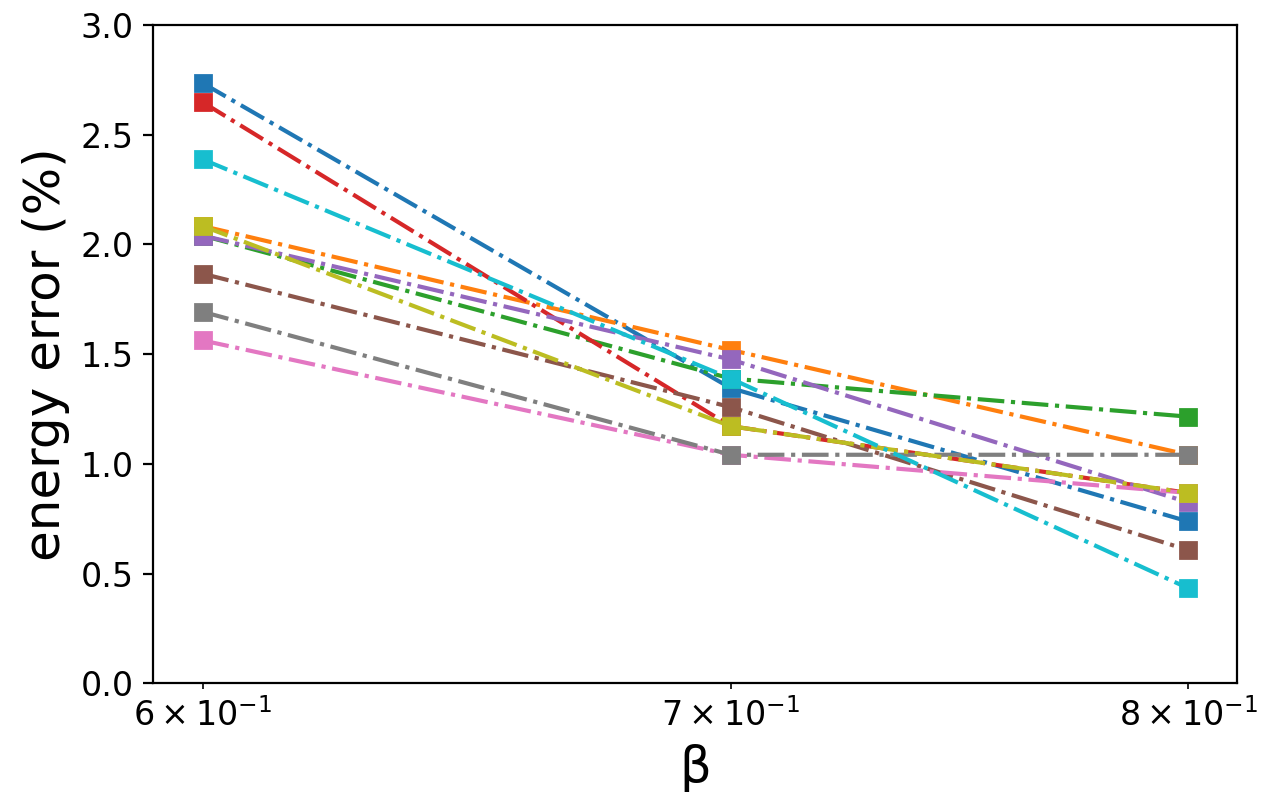}
    \caption{(color online). Square-lattice tile planting; solution quality vs. $\beta$. Ten instances from the most rugged base class ($C_2$) at $L=48$ with $\chi=4$.  Data points belonging to the same instance are connected by dash-dotted lines.  The minimum energy error over these values of $\beta$ is less than 2$\%$ for all instances.}
    \label{fig:error_C2_48x48}
\end{figure}

\begin{figure}
    \centering
    \includegraphics[width=\columnwidth]{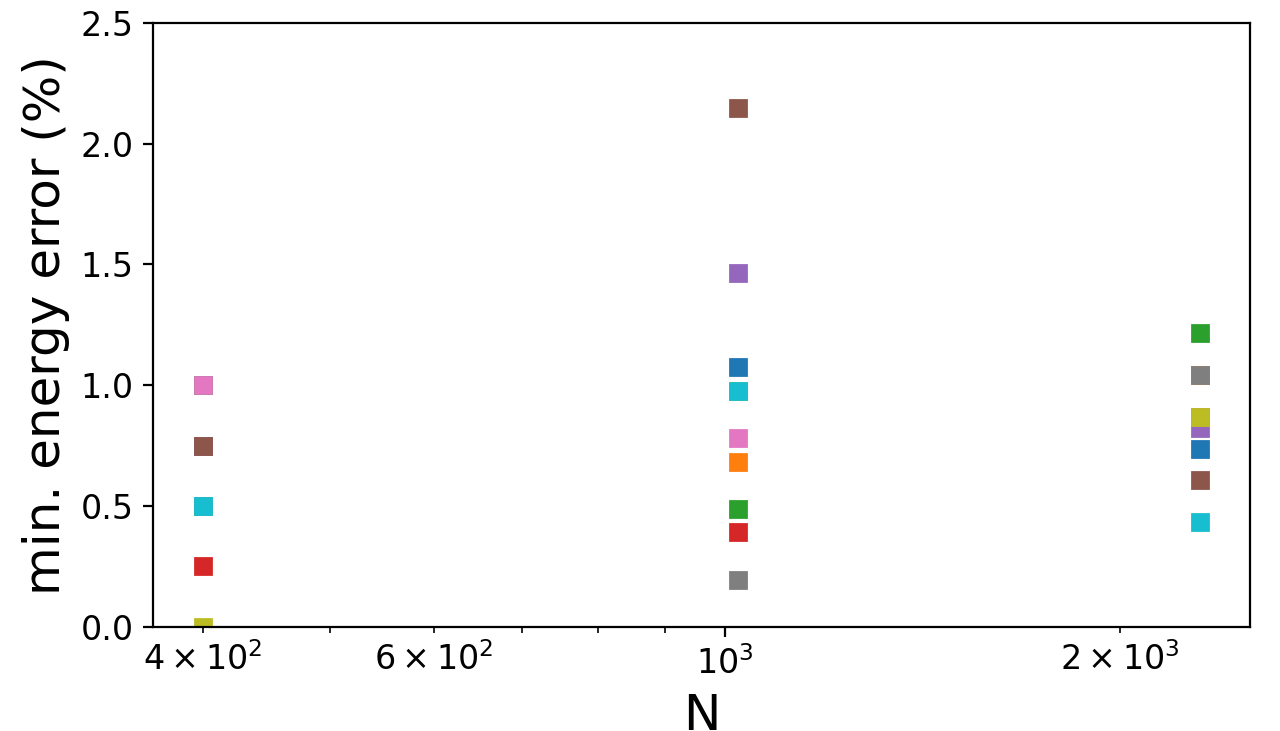}
    \caption{(color online). Square-lattice tile planting; minimum energy error vs. $N$. Ten instances in the most rugged base class ($C_2$) at each value of $N$ ($L=20, 32,$ and $48$) with $\chi=4$.  For each instance, the minimum error is determined from scanning $\beta$ (e.g., Fig. \ref{fig:error_C2_48x48}); it does not show a substantial increase with system size (in the other base classes it is zero over all instances at all tested $N$).  We demonstrate in Fig. \ref{fig:error_vs_chi_C2_L20} that the error in the $C_2$ base class at $L=20$ can be decreased by using slightly larger values of $\chi$.}
    \label{fig:minerror}
\end{figure}

\subsection{Cubic lattice tile planting}
\begin{figure*}
    \subfloat[$F_{22}$, $L=4$ ($N=64$)]{{\includegraphics[width=0.47\textwidth]{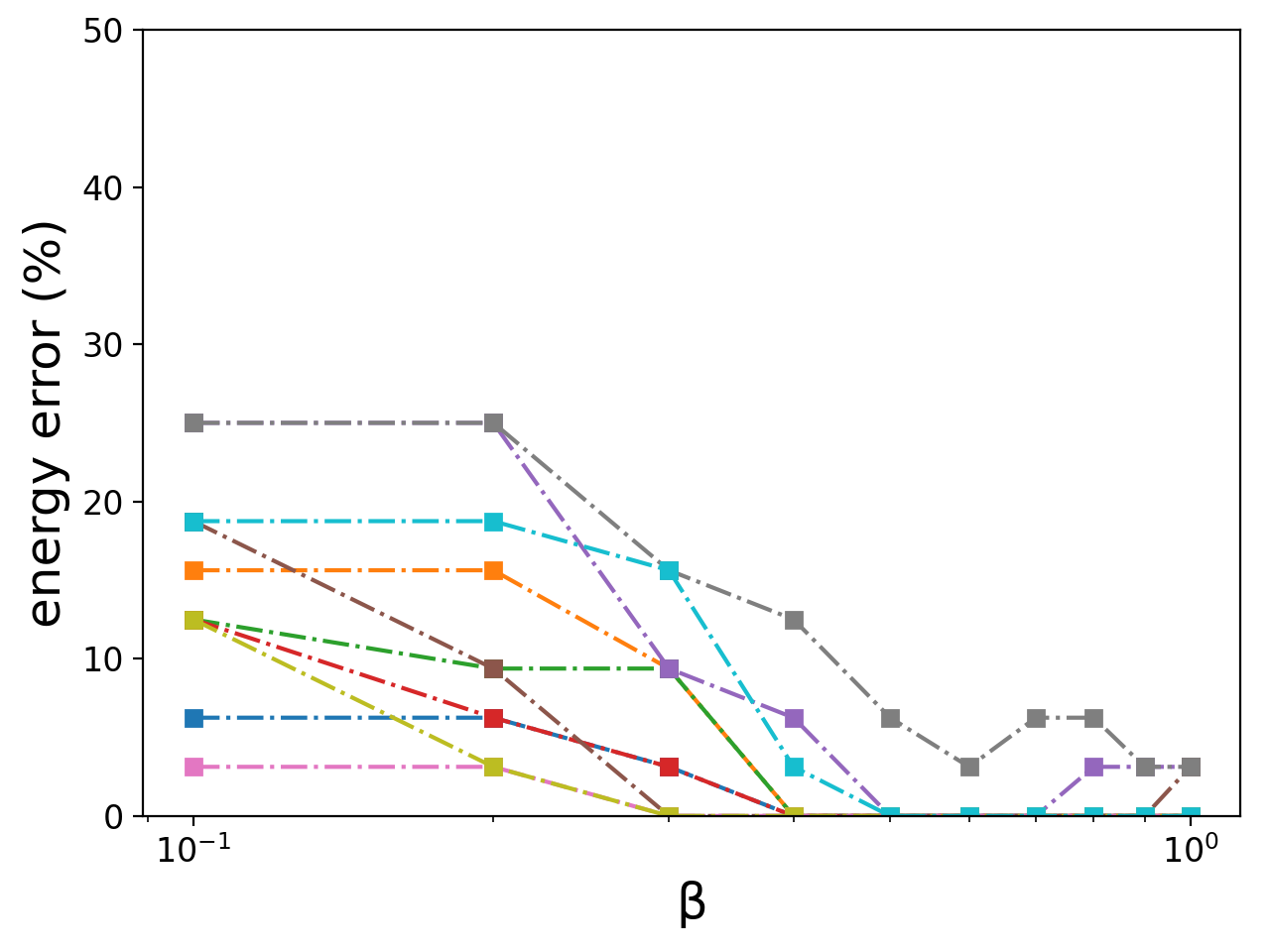} }} \hspace{0.4cm}
    \subfloat[$F_{22}$, $L=6$ ($N=216$)]{{\includegraphics[width=0.47\textwidth]{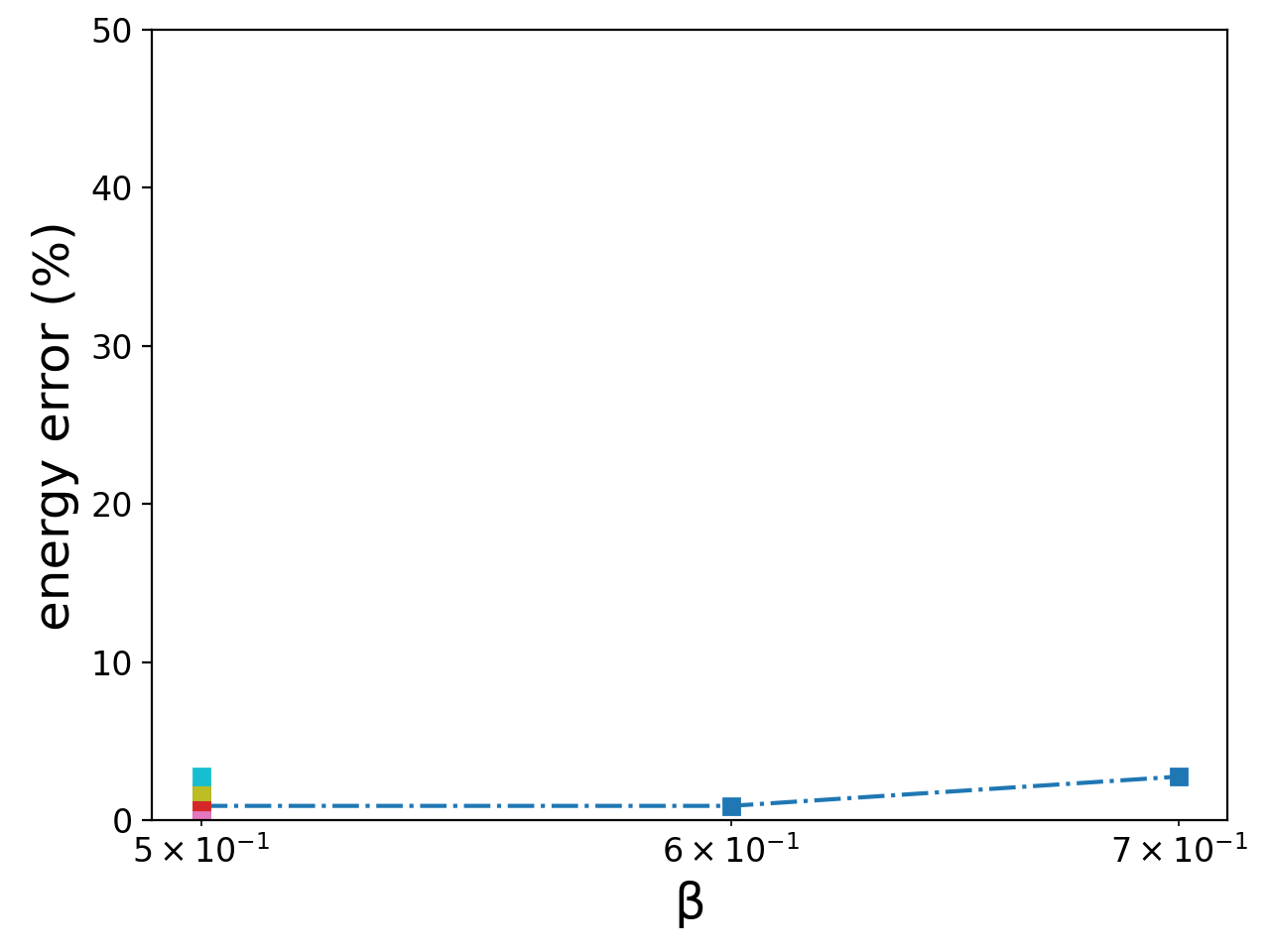} }}
    
    \subfloat[$F_{42}$, $L=4$ ($N=64$)]{{\includegraphics[width=0.47\textwidth]{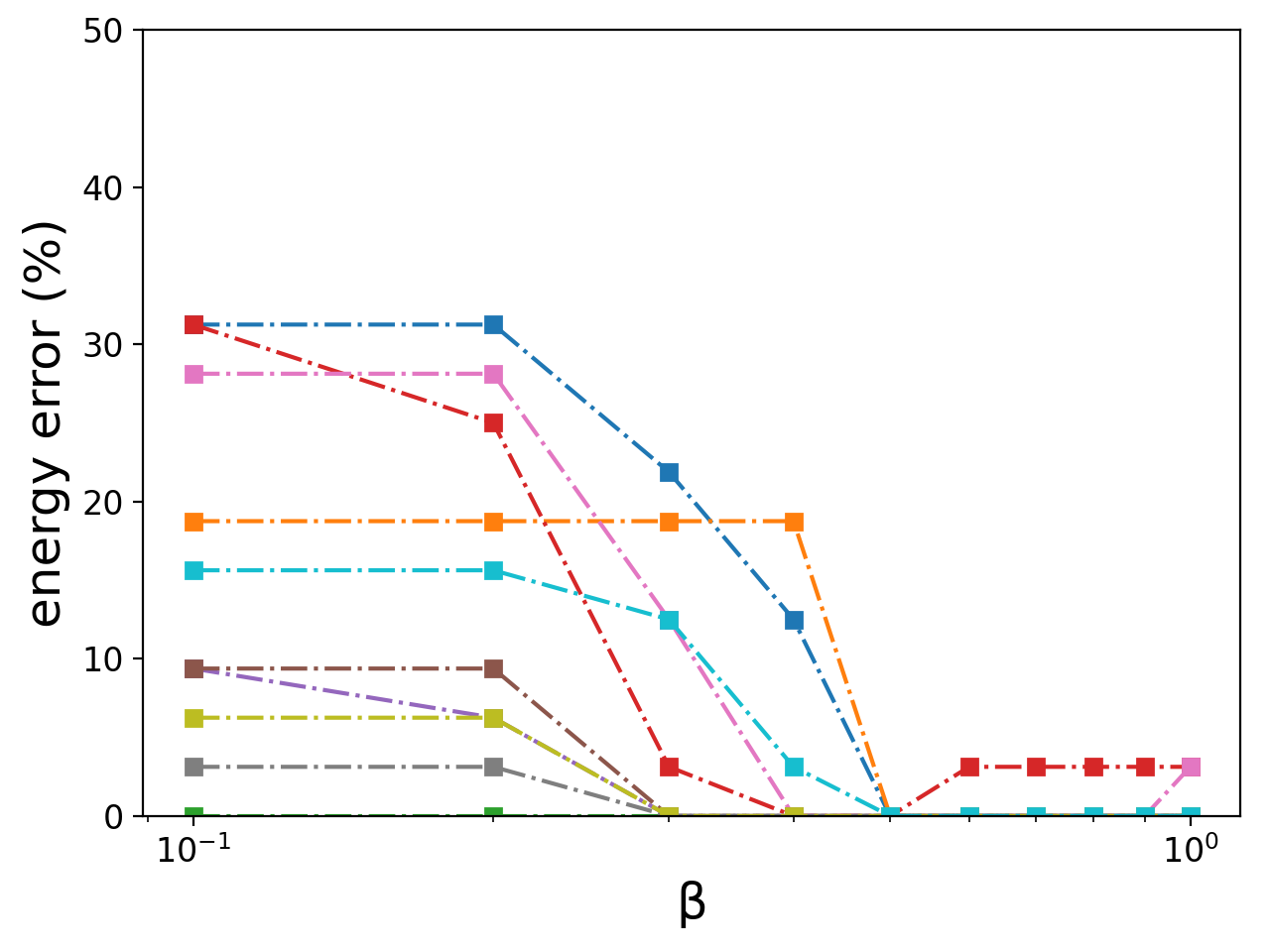} }} \hspace{0.4cm}
    \subfloat[$F_{42}$, $L=6$ ($N=216$)]{{\includegraphics[width=0.47\textwidth]{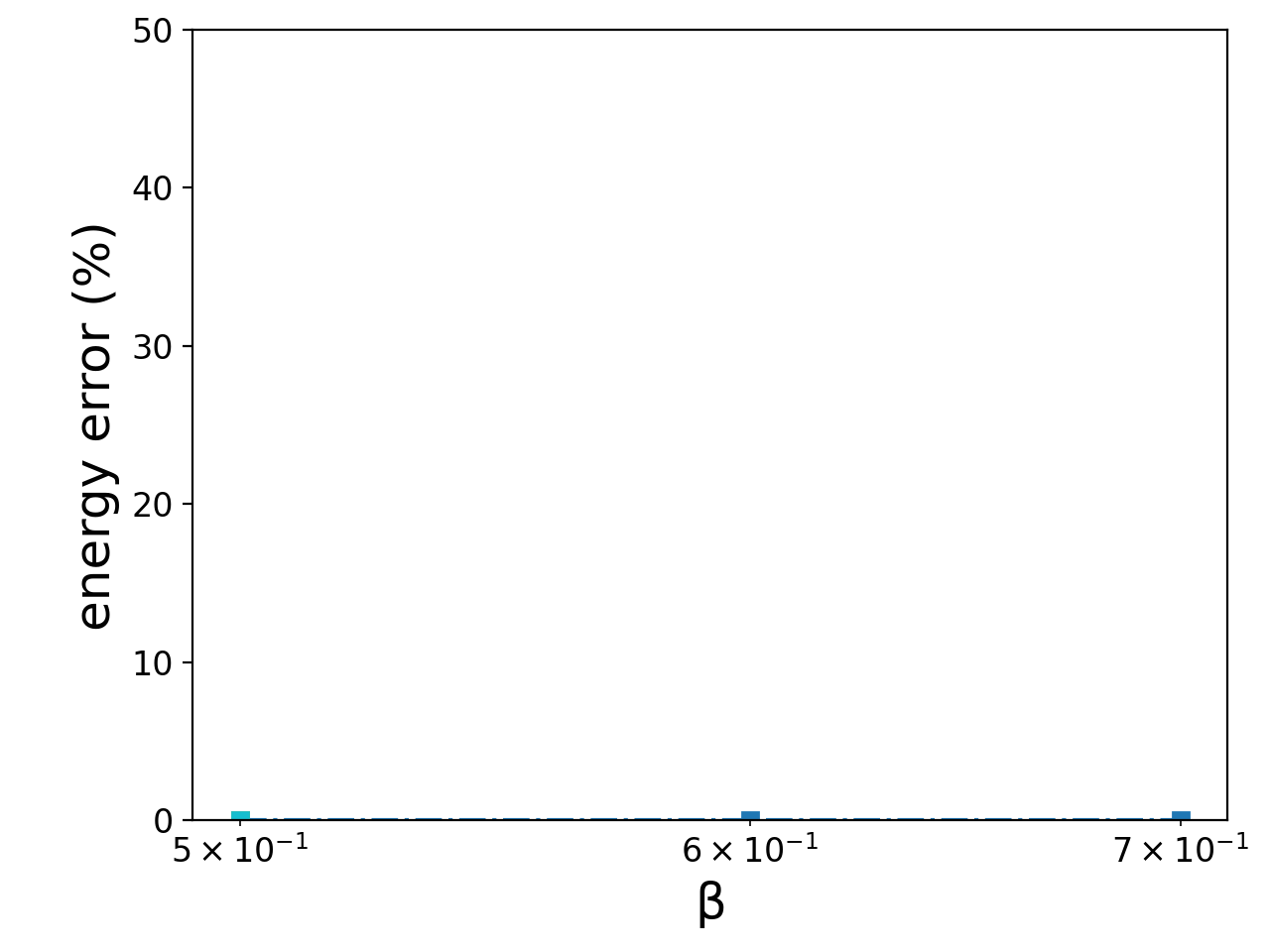} }}

    \subfloat[$F_6$, $L=4$ ($N=64$)]{{\includegraphics[width=0.47\textwidth]{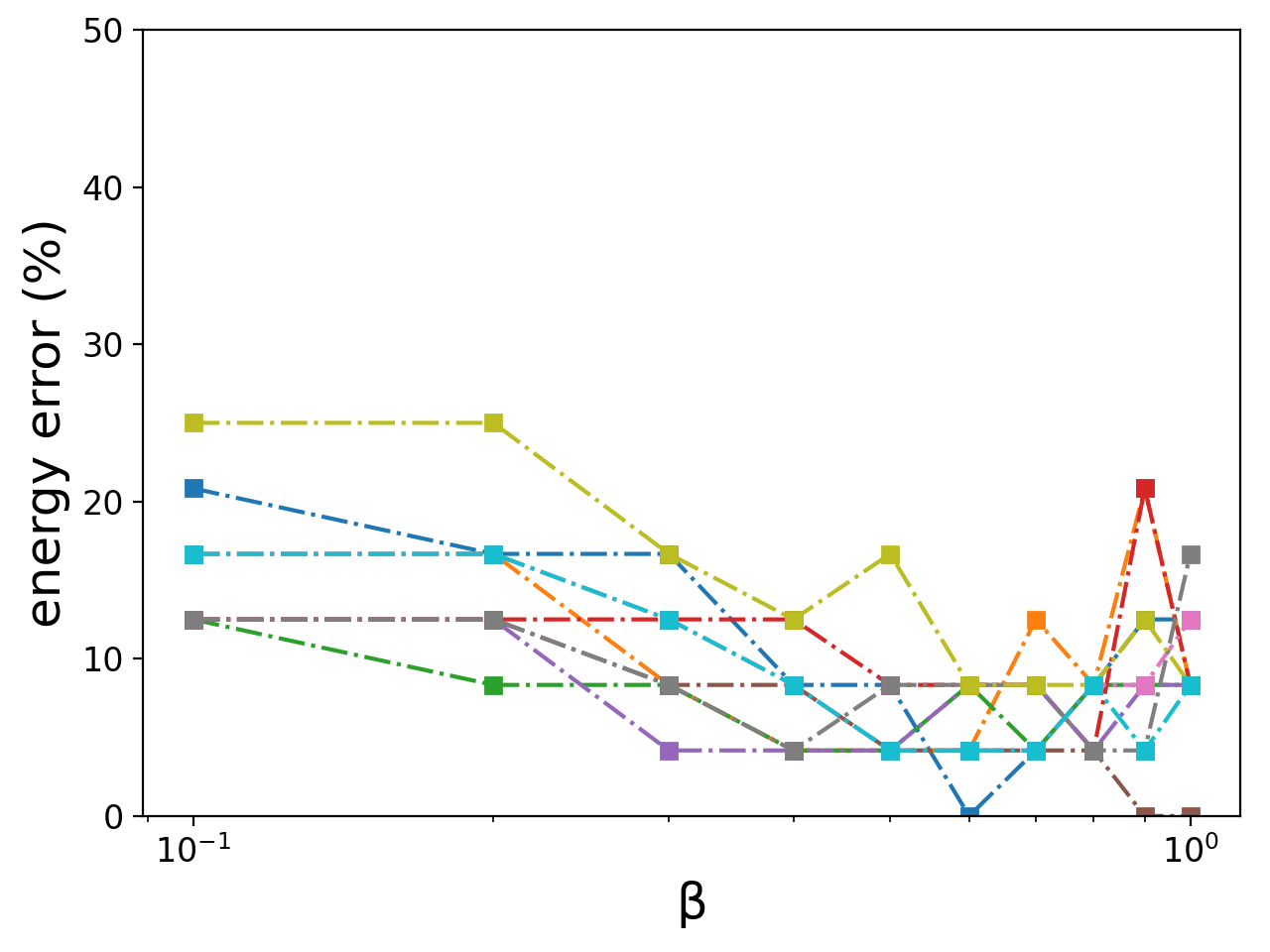} }} \hspace{0.4cm}
    \subfloat[$F_6$, $L=6$ ($N=216$)]{{\includegraphics[width=0.47\textwidth]{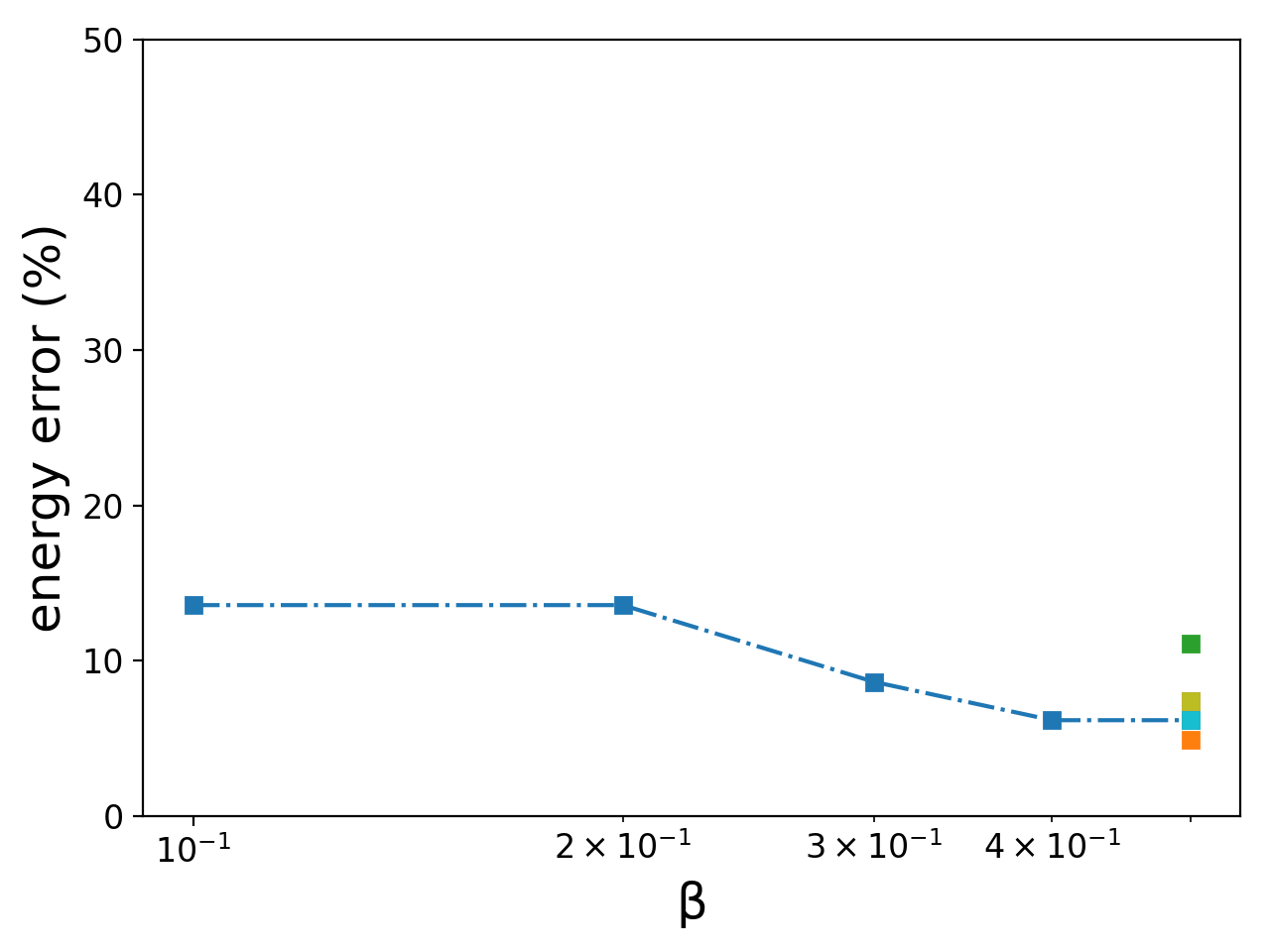} }}
    
    \caption{(color online). Cubic-lattice tile planting; error vs. $\beta$ for ten instances of each of the base classes $F_{22}$ (top row), $F_{42}$ (middle row), and $F_6$ (bottom row), all with $\chi=4$.  At $L=6$ (right column), only one of the instances are evaluated over a range of $\beta$, while the other nine instances are evaluated only at $\beta=0.5$.  Data points belonging to the same instance are connected by dash-dotted lines. The minimum energy error does not substantially increase between $L=4$ and $L=6$ within any of these base classes. We demonstrate in Fig. \ref{fig:error_vs_chi_F6_L4} that the error for some instances in the $F_6$ base class at $L=4$ can be decreased by using slightly larger values of $\chi$.}
    \label{fig:error_vs_beta_cubic_chi4}
\end{figure*}

We test our method against the $F_{22}$, $F_{42}$, and $F_6$ base classes of the cubic-lattice model individually.  Within each base class, we test ten problem instances at sizes $L=4$ ($N=L^3=64$) and $L=6$ ($N=216$).  We do not test larger sizes due to the long computation times required.  In Fig. \ref{fig:error_vs_beta_cubic_chi4} we show results for the three base classes at $L=4$ and $L=6$, all with $\chi=4$.  For the most rugged base class ($F_6$), the minimum error is less than $10\%$ at $L=4$ ($N=64$) and less than $12\%$ at $L=6$ ($N=216$).

Due to having data from only two values of $L$, we do not have a plot that meaningfully demonstrates the scaling of the time complexity, but the theoretical expectation is a quadratic scaling in $N=L^3$, as explained in Sec. \ref{sec:method}.

\begin{figure*}
    \subfloat[$\beta=0.7$]{{\includegraphics[width=\columnwidth]{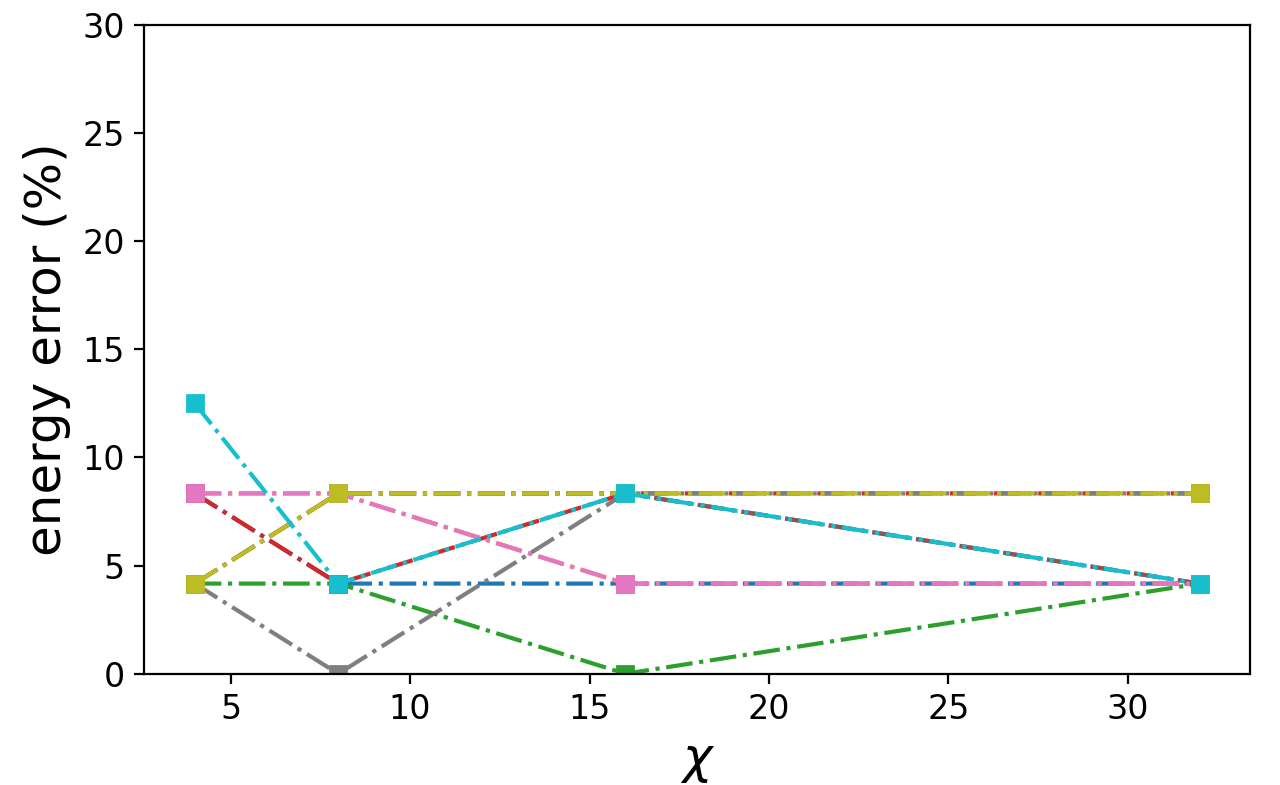} }}
    \hspace{0.4cm}
    \subfloat[$\beta=1.2$]{{\includegraphics[width=\columnwidth]{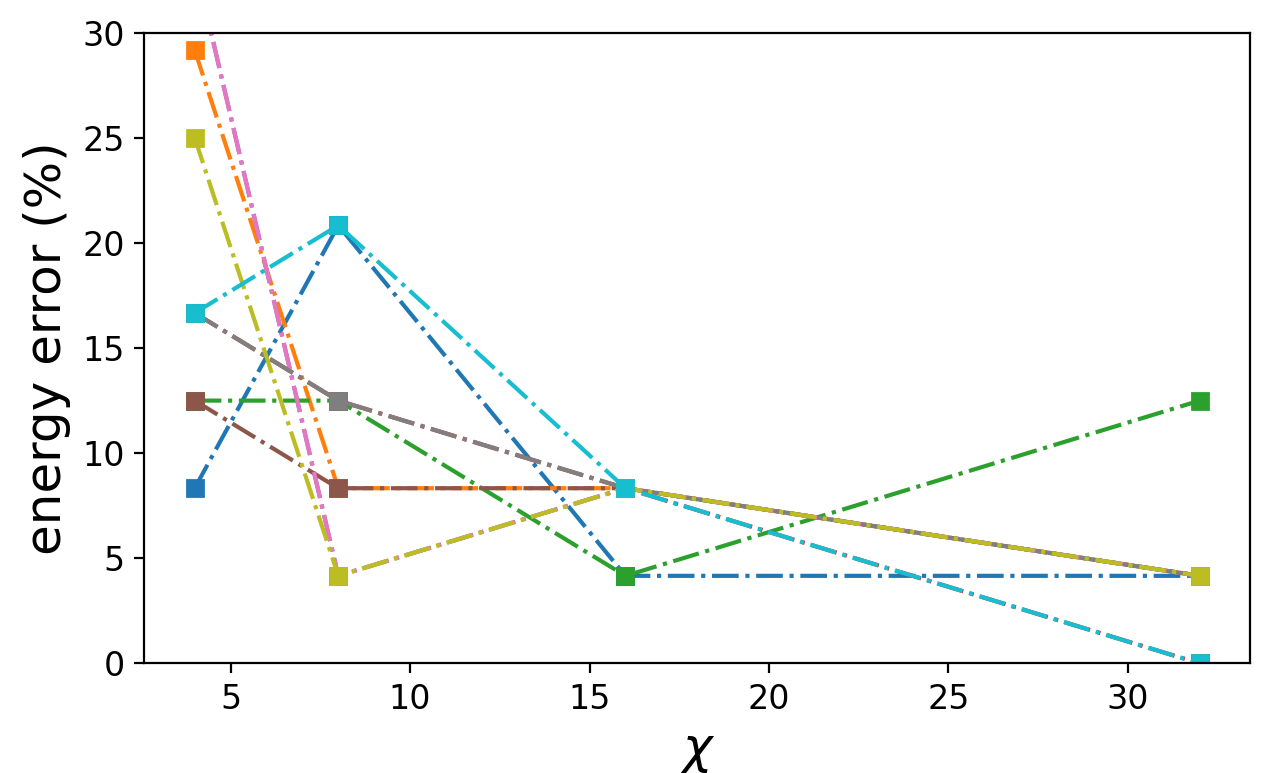} }}    
    \caption{(color online). Cubic-lattice tile planting; error vs. $\chi$ at $L=4$ for ten instances of the most rugged base class ($F_6$).  Data points belonging to the same instance are connected by dash-dotted lines.}
    \label{fig:error_vs_chi_F6_L4}
\end{figure*}

\begin{figure}
    \centering
    \includegraphics[width=\columnwidth]{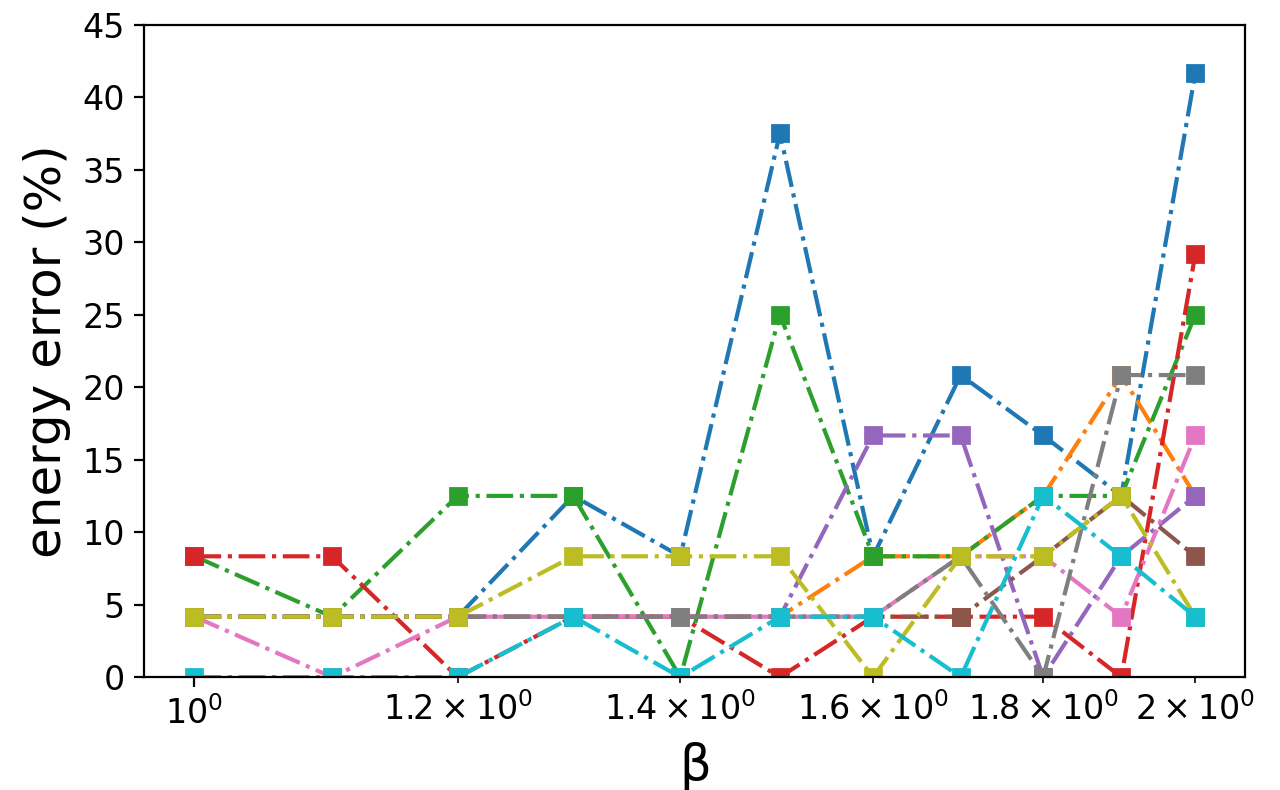}
    \caption{(color online). Cubic-lattice tile planting; error vs. $\beta$ at $L=4$ and $\chi=32$ for ten instances of the most rugged base class ($F_6$).  Data points belonging to the same instance are connected by dash-dotted lines.  The minimum average error is below 5\%.}
    \label{fig:error_vs_beta_F6_L4_chi32}
\end{figure}

\begin{figure}
    \centering
    \includegraphics[width=\columnwidth]{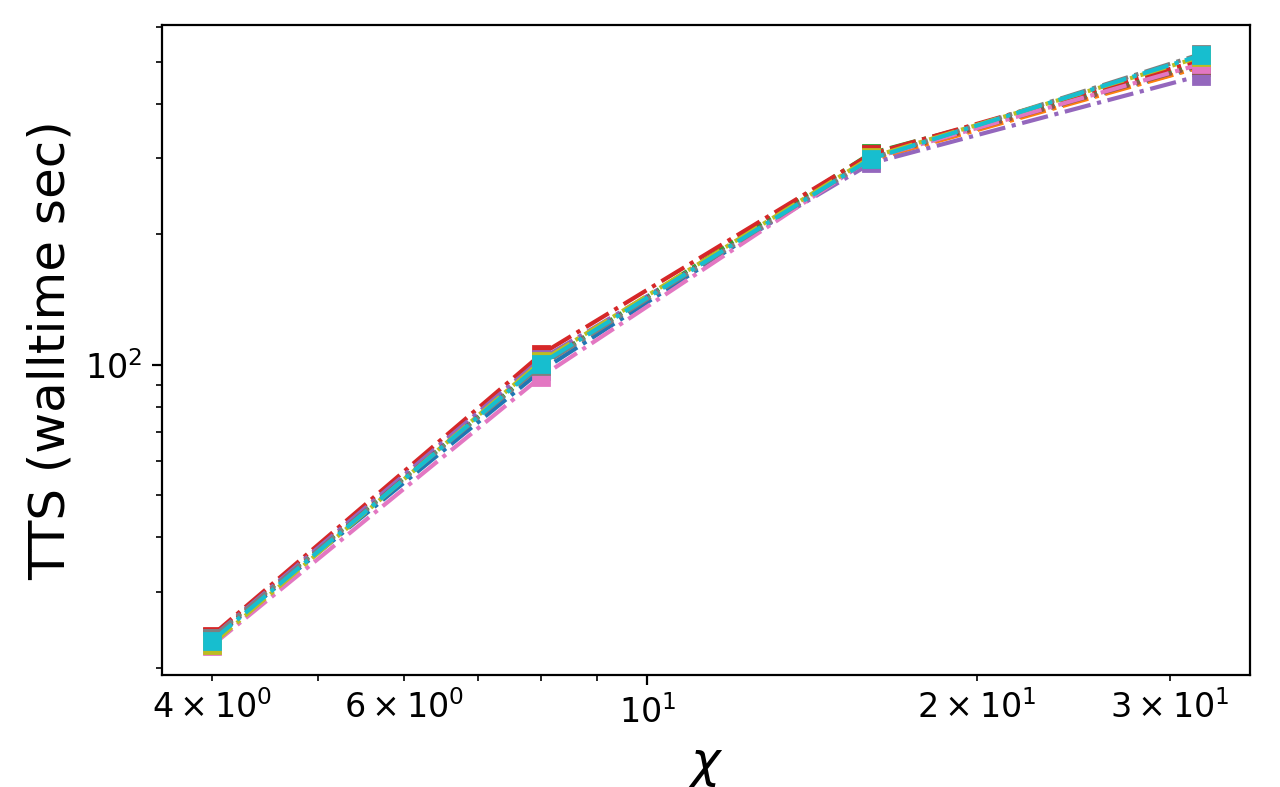}
    \caption{(color online). Cubic-lattice tile planting; TTS vs. $\chi$ at $L=4$ and $\beta=1.2$ for ten instances of the most rugged base class ($F_6$).  Data points belonging to the same instance are connected by dash-dotted lines.}
    \label{fig:TTS_vs_chi_F6_L4}
\end{figure}

We also perform some investigation of the performance in the most rugged base class ($F_6$) at $L=4$ with larger values of $\chi$.  In Fig. \ref{fig:error_vs_chi_F6_L4} we plot the energy error for all ten instances of $F_6$ at $L=4$ and two values of $\beta$ as a function of $\chi$.  This illustrates that the optimal value of $\chi$ is $\beta$-dependent, and that the error can be further decreased from what is obtained with $\chi=4$ by moderately increasing $\chi$.  In Fig. \ref{fig:error_vs_beta_F6_L4_chi32} we plot the energy error for all ten instances of $F_6$ at $L=4$ and $\chi=32$ as a function of $\beta$. Comparison with Fig. \ref{fig:error_vs_beta_cubic_chi4}(e) shows that the optimal value of $\beta$ is $\chi$-dependent.  Regarding the time cost of increasing $\chi$, in Fig. \ref{fig:TTS_vs_chi_F6_L4} we plot the TTS for all ten instances of $F_6$ at $L=4$ and $\beta=1.2$ as a function of $\chi$, revealing an approximately twenty-five times increase between $\chi=4$ and $\chi=32$.  It is unclear from the data in Fig. \ref{fig:TTS_vs_chi_F6_L4} what the asymptotic scaling of the TTS is with $\chi$, but given that the average error lies below $5\%$ for $F_6$ at $L=4$ and $\chi=32$ (see Fig. \ref{fig:error_vs_chi_F6_L4}) and that the solution quality doesn't seem to have a strong size dependence (see Fig. \ref{fig:error_vs_beta_cubic_chi4}), values of $\chi$ that are far beyond the range tested are likely not practically relevant for cubic-lattice Ising spin glasses.

\section{Summary and Discussion}

We have investigated the ability of approximate tensor network contraction with small maximum bond dimension to yield optimal or near-optimal (i.e., $\lesssim10\%$ energy error) solutions in polynomial time to at least some types of non-planar, short-range spin glass problems with rugged free energy landscapes. For planted-solution Ising spin glasses generated with tile planting, with a constant maximum bond dimension of $\chi=4$, and therefore fixed polynomial time and memory complexity, we found no substantial reduction in solution quality with increasing system size between $400$ and $2304$ spins in the toroidal-square-lattice case, and between $64$ and $216$ spins in the toroidal-cubic-lattice case.  For the most rugged subclass of instances in the toroidal-square-lattice case, the average best solution quality (over multiple instances) remained at $\sim1\%$ with increasing system size, while it remained at $\sim10\%$ for the most rugged subclass in the toroidal-cubic-lattice case.  As far as solution quality, these results are encouraging given that the sampling method that we used was quite crude; it is likely that a more sophisticated sampling procedure, such as the one in Ref. \cite{rams2021approximate}, would yield a substantial reduction in the minimum energy error for some of the problem instances without increasing $\chi$.  At the same time, we emphasize that the method here does not have a strict guarantee on solution quality (i.e., it is categorized as a \textit{heuristic} method).  Regarding system size, though the time complexity of the method is quadratic, we were unable to reach beyond $6\times6\times6$ for the cubic-lattice due to the actual simulation time required; further development of the method is needed in order to be able to use tensor-network representations of partition functions to approximately optimize the much larger cubic-lattice sizes addressable with certain other methods.

 For instance, the recent D-Wave quantum annealing experiment in Ref. \cite{king2023quantum} also yielded an approximate optimization of a cubic-lattice spin-glass model, but at a much larger size of over 5,000 spins while achieving an error of about 2$\%$ on a much shorter time scale of microseconds. A direct comparison between their results and the ones here is not possible since they did not assess the computational time complexity (they used a fixed system size), nor did they test the performance of their method against different levels of energy-landscape ruggedness, which in the literature has become synonymous with computational hardness \cite{perera2020computational, ciarella2023machine}; it is known that if a method successfully performs approximate optimization for some spin glass models at very large sizes (e.g., 8,000 spins on a cubic lattice \cite{fan2023searching}), it may still perform poorly on energy landscapes that are more rugged \cite{ciarella2023machine}.  But we can make a rough comparison between our results and those of Ref. \cite{king2023quantum} as follows: According to Ref. \cite{hamze2018near}, the most rugged subclass ($F_6$) for the cubic lattice model that we test is orders of magnitude more computationally difficult than the random $\pm J$ instances that are tested in Ref. \cite{king2023quantum}.  Considering this, our obtained error of $\sim$10\% for subclass $F_6$ (Figs. \ref{fig:error_vs_beta_cubic_chi4} (e-f)) suggests that our method (or a further development thereof) may perform competitively in terms of solution quality (i.e., relative error) for the spin glass instances of Ref. \cite{king2023quantum}.  This is further suggested in Figs. \ref{fig:error_vs_beta_cubic_chi4} (a-d), which show that when $L=6$ our method achieves an error of about 3\% for the $F_{22}$ subclass and an error of 0\% in the $F_{42}$ subclass.  Also, Fig. \ref{fig:error_vs_chi_F6_L4} (b) reveals an average error of $\sim5\%$ in the $F_6$ subclass when $L=4$ and $\chi=32$.  But as mentioned above, it is unlikely that the method here could approximately optimize a cubic lattice as large as 5,000 spins in a practical amount of time, and in that sense one can say that the D-Wave results are superior.  At the same time, the method here has a quadratic time complexity, and further development of this method may therefore eventually yield something competitive regarding reachable system sizes; the fact that small bond dimension ($\chi$) yields good results suggests that the physics is highly local, and ongoing followup work \cite{AAG} suggests that this might be leveraged to dramatically improve simulation time and thereby reach much larger system sizes.  On the other hand, there is theoretical hope that the time complexity of quantum annealing may become universally polynomial \cite{susa2018exponential, bernaschi2024quantum, zhang2024cyclic, ghosh2024exponential}, which could help quantum annealing reach still larger system sizes.  Also noteworthy is that Ref. \cite{raymond2023hybrid} demonstrates that cubic spin glasses in the form of minor embeddings that are larger than what can fit on a single quantum annealing processor can also be approximately optimized with quantum annealing by a hybrid quantum-classical approach that outperforms classical simulated annealing at a fixed system size.  Thus, the situation regarding competition between methods for optimization of cubic-lattice Ising spin glasses remains fluid.

It is also worthwhile to juxtapose our square-lattice results with those of the D-Wave quantum annealing experiment in Ref. \cite{bauza2024scaling}, which uses a quasi-two-dimensional model whose interaction graph is equivalent to a honeycomb lattice with additional non-planar, short-range bonds.  Our results (Fig. \ref{fig:minerror}) show 0\% error except in the most rugged subclass ($C_2$), where we achieve an average error of $\lesssim$1\%, and our time complexity is always quadratic.  The error-corrected quantum-annealing and parallel-tempering Monte-Carlo data in Ref. \cite{bauza2024scaling} are for error thresholds at and very close to 1\%, and always have sub-quadratic but super-linear time complexity.  The Sidon-set \cite{katzgraber2015seeking} spin glass instances that they used in Ref. \cite{bauza2024scaling} are considerably computationally harder than $\pm J$ spin glass instances, but we do not know which of the subclasses of the present square-lattice model it is closest to in terms of hardness.

Regarding the possibility of addressing more complex classical spin glasses than quasi-two-dimensional Ising and cubic-lattice Ising, the tensor-network approach investigated here is amenable to beyond-binary variables and beyond-quadratic interactions, and can easily incorporate local fields. Further, Ref. \cite{de2016simple} provides a constructive proof that square- and cubic-lattice classical Ising spin glasses faithfully capture the low-energy configuration space of \textit{any} classical spin system with only a polynomial overhead in the number of spins (see also Ref. \cite{zhang2023mapping} regarding mapping between K-SAT and the cubic-lattice classical Ising spin glass).  When dealing with such square- and cubic-lattice spin glasses that encode more general spin glasses whose interaction graphs are highly non-local, it is likely that the method investigated here would suffer from a significant bias in the sampling due to the finite bond dimension (the non-local nature of the original model that is encoded into the lattice model would likely require very large bond dimension for the present method to sample the lattice model approximately fairly).  However, some work has demonstrated that unbiased sampling of Boltzmann distributions of classical spin lattices can be achieved with tensor-network contraction by combining it with Markov Chain Monte Carlo \cite{ferris2015unbiased, huggins2017monte, frias2023collective, chen2024tensor}. In particular, Ref. \cite{chen2024tensor} demonstrated this to work well with small bond dimension for a classical square-lattice spin glass at criticality, where the correlation length is very large; it may be that the method would work well at small bond dimension also for planar or cubic-lattice spin glass models that encode spin glasses with highly non-local interaction graphs.  It is unclear, however, whether or not the time complexity would be polynomial.

Finally, short-range spin glass models are better models of real materials with quenched disorder than complete-graph models, and therefore are of relevance for condensed matter physics.  A prominent example is the classical Edwards-Anderson Ising model on the cubic lattice, the nature of the low-temperature phase of which has remained unresolved due to the rugged-energy-landscape problem \cite{charbonneau2023spin}.  Since the results in this work demonstrate that tensor-network contraction can approximately optimize rugged-energy-landscape cubic spin glasses of moderate size in polynomial time, we hope that the developments here and in related work may eventually contribute toward resolving such fundamental questions.

\section{Acknowledgments}
We acknowledge discussions with Garnet Chan and Yu Tong. AAG also acknowledges correspondence with Sam Reifenstein, Timothée Leleu, Wangwei Lan, and Marek Rams.

\bibliography{ref}

\begin{thebibliography}{59}%
\makeatletter
\providecommand \@ifxundefined [1]{%
 \@ifx{#1\undefined}
}%
\providecommand \@ifnum [1]{%
 \ifnum #1\expandafter \@firstoftwo
 \else \expandafter \@secondoftwo
 \fi
}%
\providecommand \@ifx [1]{%
 \ifx #1\expandafter \@firstoftwo
 \else \expandafter \@secondoftwo
 \fi
}%
\providecommand \natexlab [1]{#1}%
\providecommand \enquote  [1]{``#1''}%
\providecommand \bibnamefont  [1]{#1}%
\providecommand \bibfnamefont [1]{#1}%
\providecommand \citenamefont [1]{#1}%
\providecommand \href@noop [0]{\@secondoftwo}%
\providecommand \href [0]{\begingroup \@sanitize@url \@href}%
\providecommand \@href[1]{\@@startlink{#1}\@@href}%
\providecommand \@@href[1]{\endgroup#1\@@endlink}%
\providecommand \@sanitize@url [0]{\catcode `\\12\catcode `\$12\catcode
  `\&12\catcode `\#12\catcode `\^12\catcode `\_12\catcode `\%12\relax}%
\providecommand \@@startlink[1]{}%
\providecommand \@@endlink[0]{}%
\providecommand \url  [0]{\begingroup\@sanitize@url \@url }%
\providecommand \@url [1]{\endgroup\@href {#1}{\urlprefix }}%
\providecommand \urlprefix  [0]{URL }%
\providecommand \Eprint [0]{\href }%
\providecommand \doibase [0]{https://doi.org/}%
\providecommand \selectlanguage [0]{\@gobble}%
\providecommand \bibinfo  [0]{\@secondoftwo}%
\providecommand \bibfield  [0]{\@secondoftwo}%
\providecommand \translation [1]{[#1]}%
\providecommand \BibitemOpen [0]{}%
\providecommand \bibitemStop [0]{}%
\providecommand \bibitemNoStop [0]{.\EOS\space}%
\providecommand \EOS [0]{\spacefactor3000\relax}%
\providecommand \BibitemShut  [1]{\csname bibitem#1\endcsname}%
\let\auto@bib@innerbib\@empty
\bibitem [{Note1()}]{Note1}%
  \BibitemOpen
  \bibinfo {note} {The definition of ``near-optimal'' is problem-specific. For
  example, for Max-Cut problems it is in some cases desirable to have solutions
  within $\sim 12\%$ of optimal and in other cases within $\sim 6\%$ \cite
  {mohseni2022ising}.}\BibitemShut {Stop}%
\bibitem [{\citenamefont {Cai}\ \emph {et~al.}(2020)\citenamefont {Cai},
  \citenamefont {Kumar}, \citenamefont {Van~Vaerenbergh}, \citenamefont
  {Sheng}, \citenamefont {Liu}, \citenamefont {Li}, \citenamefont {Liu},
  \citenamefont {Foltin}, \citenamefont {Yu}, \citenamefont {Xia} \emph
  {et~al.}}]{cai2020power}%
  \BibitemOpen
  \bibfield  {author} {\bibinfo {author} {\bibfnamefont {F.}~\bibnamefont
  {Cai}}, \bibinfo {author} {\bibfnamefont {S.}~\bibnamefont {Kumar}}, \bibinfo
  {author} {\bibfnamefont {T.}~\bibnamefont {Van~Vaerenbergh}}, \bibinfo
  {author} {\bibfnamefont {X.}~\bibnamefont {Sheng}}, \bibinfo {author}
  {\bibfnamefont {R.}~\bibnamefont {Liu}}, \bibinfo {author} {\bibfnamefont
  {C.}~\bibnamefont {Li}}, \bibinfo {author} {\bibfnamefont {Z.}~\bibnamefont
  {Liu}}, \bibinfo {author} {\bibfnamefont {M.}~\bibnamefont {Foltin}},
  \bibinfo {author} {\bibfnamefont {S.}~\bibnamefont {Yu}}, \bibinfo {author}
  {\bibfnamefont {Q.}~\bibnamefont {Xia}}, \emph {et~al.},\ }\bibfield  {title}
  {\bibinfo {title} {Power-efficient combinatorial optimization using intrinsic
  noise in memristor hopfield neural networks},\ }\href@noop {} {\bibfield
  {journal} {\bibinfo  {journal} {Nature Electronics}\ }\textbf {\bibinfo
  {volume} {3}},\ \bibinfo {pages} {409} (\bibinfo {year} {2020})}\BibitemShut
  {NoStop}%
\bibitem [{\citenamefont {Mohseni}\ \emph {et~al.}(2022)\citenamefont
  {Mohseni}, \citenamefont {McMahon},\ and\ \citenamefont
  {Byrnes}}]{mohseni2022ising}%
  \BibitemOpen
  \bibfield  {author} {\bibinfo {author} {\bibfnamefont {N.}~\bibnamefont
  {Mohseni}}, \bibinfo {author} {\bibfnamefont {P.~L.}\ \bibnamefont
  {McMahon}},\ and\ \bibinfo {author} {\bibfnamefont {T.}~\bibnamefont
  {Byrnes}},\ }\bibfield  {title} {\bibinfo {title} {Ising machines as hardware
  solvers of combinatorial optimization problems},\ }\href@noop {} {\bibfield
  {journal} {\bibinfo  {journal} {Nature Reviews Physics}\ }\textbf {\bibinfo
  {volume} {4}},\ \bibinfo {pages} {363} (\bibinfo {year} {2022})}\BibitemShut
  {NoStop}%
\bibitem [{\citenamefont {Nguyen}\ \emph {et~al.}(2024)\citenamefont {Nguyen},
  \citenamefont {Miri}, \citenamefont {Rupert}, \citenamefont {Dyk},
  \citenamefont {Wu}, \citenamefont {Vrahoretis}, \citenamefont {Huang},
  \citenamefont {Begliarbekov}, \citenamefont {Chancellor}, \citenamefont
  {Chukwu} \emph {et~al.}}]{nguyen2024entropy}%
  \BibitemOpen
  \bibfield  {author} {\bibinfo {author} {\bibfnamefont {L.}~\bibnamefont
  {Nguyen}}, \bibinfo {author} {\bibfnamefont {M.-A.}\ \bibnamefont {Miri}},
  \bibinfo {author} {\bibfnamefont {R.~J.}\ \bibnamefont {Rupert}}, \bibinfo
  {author} {\bibfnamefont {W.}~\bibnamefont {Dyk}}, \bibinfo {author}
  {\bibfnamefont {S.}~\bibnamefont {Wu}}, \bibinfo {author} {\bibfnamefont
  {N.}~\bibnamefont {Vrahoretis}}, \bibinfo {author} {\bibfnamefont
  {I.}~\bibnamefont {Huang}}, \bibinfo {author} {\bibfnamefont
  {M.}~\bibnamefont {Begliarbekov}}, \bibinfo {author} {\bibfnamefont
  {N.}~\bibnamefont {Chancellor}}, \bibinfo {author} {\bibfnamefont
  {U.}~\bibnamefont {Chukwu}}, \emph {et~al.},\ }\bibfield  {title} {\bibinfo
  {title} {Entropy computing: A paradigm for optimization in an open quantum
  system},\ }\href@noop {} {\bibfield  {journal} {\bibinfo  {journal} {arXiv
  preprint arXiv:2407.04512}\ } (\bibinfo {year} {2024})}\BibitemShut {NoStop}%
\bibitem [{\citenamefont {Zhang}\ \emph
  {et~al.}(2024{\natexlab{a}})\citenamefont {Zhang}, \citenamefont {Tao},
  \citenamefont {Liu}, \citenamefont {Grimaldi}, \citenamefont {Raimondo},
  \citenamefont {Jimenez}, \citenamefont {Avedillo}, \citenamefont {Nunez},
  \citenamefont {Linares-Barranco}, \citenamefont {Serrano-Gotarredona} \emph
  {et~al.}}]{zhang2024review}%
  \BibitemOpen
  \bibfield  {author} {\bibinfo {author} {\bibfnamefont {T.}~\bibnamefont
  {Zhang}}, \bibinfo {author} {\bibfnamefont {Q.}~\bibnamefont {Tao}}, \bibinfo
  {author} {\bibfnamefont {B.}~\bibnamefont {Liu}}, \bibinfo {author}
  {\bibfnamefont {A.}~\bibnamefont {Grimaldi}}, \bibinfo {author}
  {\bibfnamefont {E.}~\bibnamefont {Raimondo}}, \bibinfo {author}
  {\bibfnamefont {M.}~\bibnamefont {Jimenez}}, \bibinfo {author} {\bibfnamefont
  {M.~J.}\ \bibnamefont {Avedillo}}, \bibinfo {author} {\bibfnamefont
  {J.}~\bibnamefont {Nunez}}, \bibinfo {author} {\bibfnamefont
  {B.}~\bibnamefont {Linares-Barranco}}, \bibinfo {author} {\bibfnamefont
  {T.}~\bibnamefont {Serrano-Gotarredona}}, \emph {et~al.},\ }\bibfield
  {title} {\bibinfo {title} {A review of ising machines implemented in
  conventional and emerging technologies},\ }\href@noop {} {\bibfield
  {journal} {\bibinfo  {journal} {IEEE Transactions on Nanotechnology}\ }
  (\bibinfo {year} {2024}{\natexlab{a}})}\BibitemShut {NoStop}%
\bibitem [{\citenamefont {King}\ \emph {et~al.}(2024)\citenamefont {King},
  \citenamefont {Nocera}, \citenamefont {Rams}, \citenamefont {Dziarmaga},
  \citenamefont {Wiersema}, \citenamefont {Bernoudy}, \citenamefont {Raymond},
  \citenamefont {Kaushal}, \citenamefont {Heinsdorf}, \citenamefont {Harris}
  \emph {et~al.}}]{king2024computational}%
  \BibitemOpen
  \bibfield  {author} {\bibinfo {author} {\bibfnamefont {A.~D.}\ \bibnamefont
  {King}}, \bibinfo {author} {\bibfnamefont {A.}~\bibnamefont {Nocera}},
  \bibinfo {author} {\bibfnamefont {M.~M.}\ \bibnamefont {Rams}}, \bibinfo
  {author} {\bibfnamefont {J.}~\bibnamefont {Dziarmaga}}, \bibinfo {author}
  {\bibfnamefont {R.}~\bibnamefont {Wiersema}}, \bibinfo {author}
  {\bibfnamefont {W.}~\bibnamefont {Bernoudy}}, \bibinfo {author}
  {\bibfnamefont {J.}~\bibnamefont {Raymond}}, \bibinfo {author} {\bibfnamefont
  {N.}~\bibnamefont {Kaushal}}, \bibinfo {author} {\bibfnamefont
  {N.}~\bibnamefont {Heinsdorf}}, \bibinfo {author} {\bibfnamefont
  {R.}~\bibnamefont {Harris}}, \emph {et~al.},\ }\bibfield  {title} {\bibinfo
  {title} {Computational supremacy in quantum simulation},\ }\href@noop {}
  {\bibfield  {journal} {\bibinfo  {journal} {arXiv preprint arXiv:2403.00910}\
  } (\bibinfo {year} {2024})}\BibitemShut {NoStop}%
\bibitem [{\citenamefont {Traversa}\ and\ \citenamefont
  {Di~Ventra}(2017)}]{traversa2017polynomial}%
  \BibitemOpen
  \bibfield  {author} {\bibinfo {author} {\bibfnamefont {F.~L.}\ \bibnamefont
  {Traversa}}\ and\ \bibinfo {author} {\bibfnamefont {M.}~\bibnamefont
  {Di~Ventra}},\ }\bibfield  {title} {\bibinfo {title} {Polynomial-time
  solution of prime factorization and np-complete problems with digital
  memcomputing machines},\ }\href@noop {} {\bibfield  {journal} {\bibinfo
  {journal} {Chaos: An Interdisciplinary Journal of Nonlinear Science}\
  }\textbf {\bibinfo {volume} {27}} (\bibinfo {year} {2017})}\BibitemShut
  {NoStop}%
\bibitem [{\citenamefont {Bernaschi}\ \emph {et~al.}(2024)\citenamefont
  {Bernaschi}, \citenamefont {Gonz{\'a}lez-Adalid~Pemart{\'\i}n}, \citenamefont
  {Mart{\'\i}n-Mayor},\ and\ \citenamefont {Parisi}}]{bernaschi2024quantum}%
  \BibitemOpen
  \bibfield  {author} {\bibinfo {author} {\bibfnamefont {M.}~\bibnamefont
  {Bernaschi}}, \bibinfo {author} {\bibfnamefont {I.}~\bibnamefont
  {Gonz{\'a}lez-Adalid~Pemart{\'\i}n}}, \bibinfo {author} {\bibfnamefont
  {V.}~\bibnamefont {Mart{\'\i}n-Mayor}},\ and\ \bibinfo {author}
  {\bibfnamefont {G.}~\bibnamefont {Parisi}},\ }\bibfield  {title} {\bibinfo
  {title} {The quantum transition of the two-dimensional ising spin glass},\
  }\href@noop {} {\bibfield  {journal} {\bibinfo  {journal} {Nature}\ }\textbf
  {\bibinfo {volume} {631}},\ \bibinfo {pages} {749} (\bibinfo {year}
  {2024})}\BibitemShut {NoStop}%
\bibitem [{\citenamefont {Zhang}\ \emph
  {et~al.}(2024{\natexlab{b}})\citenamefont {Zhang}, \citenamefont {Boothby},\
  and\ \citenamefont {Kamenev}}]{zhang2024cyclic}%
  \BibitemOpen
  \bibfield  {author} {\bibinfo {author} {\bibfnamefont {H.}~\bibnamefont
  {Zhang}}, \bibinfo {author} {\bibfnamefont {K.}~\bibnamefont {Boothby}},\
  and\ \bibinfo {author} {\bibfnamefont {A.}~\bibnamefont {Kamenev}},\
  }\bibfield  {title} {\bibinfo {title} {Cyclic quantum annealing: Searching
  for deep low-energy states in 5000-qubit spin glass},\ }\href@noop {}
  {\bibfield  {journal} {\bibinfo  {journal} {arXiv preprint arXiv:2403.01034}\
  } (\bibinfo {year} {2024}{\natexlab{b}})}\BibitemShut {NoStop}%
\bibitem [{\citenamefont {Ghosh}\ \emph {et~al.}(2024)\citenamefont {Ghosh},
  \citenamefont {Nutricati}, \citenamefont {Feinstein}, \citenamefont
  {Warburton},\ and\ \citenamefont {Bose}}]{ghosh2024exponential}%
  \BibitemOpen
  \bibfield  {author} {\bibinfo {author} {\bibfnamefont {R.}~\bibnamefont
  {Ghosh}}, \bibinfo {author} {\bibfnamefont {L.~A.}\ \bibnamefont
  {Nutricati}}, \bibinfo {author} {\bibfnamefont {N.}~\bibnamefont
  {Feinstein}}, \bibinfo {author} {\bibfnamefont {P.}~\bibnamefont
  {Warburton}},\ and\ \bibinfo {author} {\bibfnamefont {S.}~\bibnamefont
  {Bose}},\ }\bibfield  {title} {\bibinfo {title} {Exponential speed-up of
  quantum annealing via n-local catalysts},\ }\href@noop {} {\bibfield
  {journal} {\bibinfo  {journal} {arXiv preprint arXiv:2409.13029}\ } (\bibinfo
  {year} {2024})}\BibitemShut {NoStop}%
\bibitem [{\citenamefont {King}\ \emph {et~al.}(2023)\citenamefont {King},
  \citenamefont {Raymond}, \citenamefont {Lanting}, \citenamefont {Harris},
  \citenamefont {Zucca}, \citenamefont {Altomare}, \citenamefont {Berkley},
  \citenamefont {Boothby}, \citenamefont {Ejtemaee}, \citenamefont {Enderud}
  \emph {et~al.}}]{king2023quantum}%
  \BibitemOpen
  \bibfield  {author} {\bibinfo {author} {\bibfnamefont {A.~D.}\ \bibnamefont
  {King}}, \bibinfo {author} {\bibfnamefont {J.}~\bibnamefont {Raymond}},
  \bibinfo {author} {\bibfnamefont {T.}~\bibnamefont {Lanting}}, \bibinfo
  {author} {\bibfnamefont {R.}~\bibnamefont {Harris}}, \bibinfo {author}
  {\bibfnamefont {A.}~\bibnamefont {Zucca}}, \bibinfo {author} {\bibfnamefont
  {F.}~\bibnamefont {Altomare}}, \bibinfo {author} {\bibfnamefont {A.~J.}\
  \bibnamefont {Berkley}}, \bibinfo {author} {\bibfnamefont {K.}~\bibnamefont
  {Boothby}}, \bibinfo {author} {\bibfnamefont {S.}~\bibnamefont {Ejtemaee}},
  \bibinfo {author} {\bibfnamefont {C.}~\bibnamefont {Enderud}}, \emph
  {et~al.},\ }\bibfield  {title} {\bibinfo {title} {Quantum critical dynamics
  in a 5,000-qubit programmable spin glass},\ }\href@noop {} {\bibfield
  {journal} {\bibinfo  {journal} {Nature}\ }\textbf {\bibinfo {volume} {617}},\
  \bibinfo {pages} {61} (\bibinfo {year} {2023})}\BibitemShut {NoStop}%
\bibitem [{\citenamefont {Bauza}\ and\ \citenamefont
  {Lidar}(2024)}]{bauza2024scaling}%
  \BibitemOpen
  \bibfield  {author} {\bibinfo {author} {\bibfnamefont {H.~M.}\ \bibnamefont
  {Bauza}}\ and\ \bibinfo {author} {\bibfnamefont {D.~A.}\ \bibnamefont
  {Lidar}},\ }\bibfield  {title} {\bibinfo {title} {Scaling advantage in
  approximate optimization with quantum annealing},\ }\href@noop {} {\bibfield
  {journal} {\bibinfo  {journal} {arXiv preprint arXiv:2401.07184}\ } (\bibinfo
  {year} {2024})}\BibitemShut {NoStop}%
\bibitem [{\citenamefont {Zhu}\ \emph {et~al.}(2015)\citenamefont {Zhu},
  \citenamefont {Ochoa},\ and\ \citenamefont {Katzgraber}}]{zhu2015efficient}%
  \BibitemOpen
  \bibfield  {author} {\bibinfo {author} {\bibfnamefont {Z.}~\bibnamefont
  {Zhu}}, \bibinfo {author} {\bibfnamefont {A.~J.}\ \bibnamefont {Ochoa}},\
  and\ \bibinfo {author} {\bibfnamefont {H.~G.}\ \bibnamefont {Katzgraber}},\
  }\bibfield  {title} {\bibinfo {title} {Efficient cluster algorithm for spin
  glasses in any space dimension},\ }\href@noop {} {\bibfield  {journal}
  {\bibinfo  {journal} {Physical review letters}\ }\textbf {\bibinfo {volume}
  {115}},\ \bibinfo {pages} {077201} (\bibinfo {year} {2015})}\BibitemShut
  {NoStop}%
\bibitem [{\citenamefont {K{\"o}nz}\ \emph {et~al.}(2021)\citenamefont
  {K{\"o}nz}, \citenamefont {Lechner}, \citenamefont {Katzgraber},\ and\
  \citenamefont {Troyer}}]{konz2021embedding}%
  \BibitemOpen
  \bibfield  {author} {\bibinfo {author} {\bibfnamefont {M.~S.}\ \bibnamefont
  {K{\"o}nz}}, \bibinfo {author} {\bibfnamefont {W.}~\bibnamefont {Lechner}},
  \bibinfo {author} {\bibfnamefont {H.~G.}\ \bibnamefont {Katzgraber}},\ and\
  \bibinfo {author} {\bibfnamefont {M.}~\bibnamefont {Troyer}},\ }\bibfield
  {title} {\bibinfo {title} {Embedding overhead scaling of optimization
  problems in quantum annealing},\ }\href@noop {} {\bibfield  {journal}
  {\bibinfo  {journal} {PRX Quantum}\ }\textbf {\bibinfo {volume} {2}},\
  \bibinfo {pages} {040322} (\bibinfo {year} {2021})}\BibitemShut {NoStop}%
\bibitem [{\citenamefont {Hamerly}\ \emph {et~al.}(2019)\citenamefont
  {Hamerly}, \citenamefont {Inagaki}, \citenamefont {McMahon}, \citenamefont
  {Venturelli}, \citenamefont {Marandi}, \citenamefont {Onodera}, \citenamefont
  {Ng}, \citenamefont {Langrock}, \citenamefont {Inaba}, \citenamefont {Honjo}
  \emph {et~al.}}]{hamerly2019experimental}%
  \BibitemOpen
  \bibfield  {author} {\bibinfo {author} {\bibfnamefont {R.}~\bibnamefont
  {Hamerly}}, \bibinfo {author} {\bibfnamefont {T.}~\bibnamefont {Inagaki}},
  \bibinfo {author} {\bibfnamefont {P.~L.}\ \bibnamefont {McMahon}}, \bibinfo
  {author} {\bibfnamefont {D.}~\bibnamefont {Venturelli}}, \bibinfo {author}
  {\bibfnamefont {A.}~\bibnamefont {Marandi}}, \bibinfo {author} {\bibfnamefont
  {T.}~\bibnamefont {Onodera}}, \bibinfo {author} {\bibfnamefont
  {E.}~\bibnamefont {Ng}}, \bibinfo {author} {\bibfnamefont {C.}~\bibnamefont
  {Langrock}}, \bibinfo {author} {\bibfnamefont {K.}~\bibnamefont {Inaba}},
  \bibinfo {author} {\bibfnamefont {T.}~\bibnamefont {Honjo}}, \emph {et~al.},\
  }\bibfield  {title} {\bibinfo {title} {Experimental investigation of
  performance differences between coherent ising machines and a quantum
  annealer},\ }\href@noop {} {\bibfield  {journal} {\bibinfo  {journal}
  {Science advances}\ }\textbf {\bibinfo {volume} {5}},\ \bibinfo {pages}
  {eaau0823} (\bibinfo {year} {2019})}\BibitemShut {NoStop}%
\bibitem [{\citenamefont {Harrigan}\ \emph {et~al.}(2021)\citenamefont
  {Harrigan}, \citenamefont {Sung}, \citenamefont {Neeley}, \citenamefont
  {Satzinger}, \citenamefont {Arute}, \citenamefont {Arya}, \citenamefont
  {Atalaya}, \citenamefont {Bardin}, \citenamefont {Barends}, \citenamefont
  {Boixo} \emph {et~al.}}]{harrigan2021quantum}%
  \BibitemOpen
  \bibfield  {author} {\bibinfo {author} {\bibfnamefont {M.~P.}\ \bibnamefont
  {Harrigan}}, \bibinfo {author} {\bibfnamefont {K.~J.}\ \bibnamefont {Sung}},
  \bibinfo {author} {\bibfnamefont {M.}~\bibnamefont {Neeley}}, \bibinfo
  {author} {\bibfnamefont {K.~J.}\ \bibnamefont {Satzinger}}, \bibinfo {author}
  {\bibfnamefont {F.}~\bibnamefont {Arute}}, \bibinfo {author} {\bibfnamefont
  {K.}~\bibnamefont {Arya}}, \bibinfo {author} {\bibfnamefont {J.}~\bibnamefont
  {Atalaya}}, \bibinfo {author} {\bibfnamefont {J.~C.}\ \bibnamefont {Bardin}},
  \bibinfo {author} {\bibfnamefont {R.}~\bibnamefont {Barends}}, \bibinfo
  {author} {\bibfnamefont {S.}~\bibnamefont {Boixo}}, \emph {et~al.},\
  }\bibfield  {title} {\bibinfo {title} {Quantum approximate optimization of
  non-planar graph problems on a planar superconducting processor},\
  }\href@noop {} {\bibfield  {journal} {\bibinfo  {journal} {Nature Physics}\
  }\textbf {\bibinfo {volume} {17}},\ \bibinfo {pages} {332} (\bibinfo {year}
  {2021})}\BibitemShut {NoStop}%
\bibitem [{\citenamefont {Krauth}(2006)}]{krauth2006statistical}%
  \BibitemOpen
  \bibfield  {author} {\bibinfo {author} {\bibfnamefont {W.}~\bibnamefont
  {Krauth}},\ }\href@noop {} {\emph {\bibinfo {title} {Statistical mechanics:
  algorithms and computations}}},\ Vol.~\bibinfo {volume} {13}\ (\bibinfo
  {publisher} {OUP Oxford},\ \bibinfo {year} {2006})\BibitemShut {NoStop}%
\bibitem [{\citenamefont {Machta}(2009)}]{machta2009strengths}%
  \BibitemOpen
  \bibfield  {author} {\bibinfo {author} {\bibfnamefont {J.}~\bibnamefont
  {Machta}},\ }\bibfield  {title} {\bibinfo {title} {Strengths and weaknesses
  of parallel tempering},\ }\href@noop {} {\bibfield  {journal} {\bibinfo
  {journal} {Physical Review E}\ }\textbf {\bibinfo {volume} {80}},\ \bibinfo
  {pages} {056706} (\bibinfo {year} {2009})}\BibitemShut {NoStop}%
\bibitem [{\citenamefont {Ciarella}\ \emph {et~al.}(2023)\citenamefont
  {Ciarella}, \citenamefont {Trinquier}, \citenamefont {Weigt},\ and\
  \citenamefont {Zamponi}}]{ciarella2023machine}%
  \BibitemOpen
  \bibfield  {author} {\bibinfo {author} {\bibfnamefont {S.}~\bibnamefont
  {Ciarella}}, \bibinfo {author} {\bibfnamefont {J.}~\bibnamefont {Trinquier}},
  \bibinfo {author} {\bibfnamefont {M.}~\bibnamefont {Weigt}},\ and\ \bibinfo
  {author} {\bibfnamefont {F.}~\bibnamefont {Zamponi}},\ }\bibfield  {title}
  {\bibinfo {title} {Machine-learning-assisted monte carlo fails at sampling
  computationally hard problems},\ }\href@noop {} {\bibfield  {journal}
  {\bibinfo  {journal} {Machine Learning: Science and Technology}\ }\textbf
  {\bibinfo {volume} {4}},\ \bibinfo {pages} {010501} (\bibinfo {year}
  {2023})}\BibitemShut {NoStop}%
\bibitem [{\citenamefont {Montanari}\ and\ \citenamefont
  {Semerjian}(2006)}]{montanari2006rigorous}%
  \BibitemOpen
  \bibfield  {author} {\bibinfo {author} {\bibfnamefont {A.}~\bibnamefont
  {Montanari}}\ and\ \bibinfo {author} {\bibfnamefont {G.}~\bibnamefont
  {Semerjian}},\ }\bibfield  {title} {\bibinfo {title} {Rigorous inequalities
  between length and time scales in glassy systems},\ }\href@noop {} {\bibfield
   {journal} {\bibinfo  {journal} {Journal of Statistical Physics}\ }\textbf
  {\bibinfo {volume} {125}},\ \bibinfo {pages} {23} (\bibinfo {year}
  {2006})}\BibitemShut {NoStop}%
\bibitem [{\citenamefont {Jaum{\`a}}\ \emph {et~al.}(2024)\citenamefont
  {Jaum{\`a}}, \citenamefont {Garc{\'\i}a-Ripoll},\ and\ \citenamefont
  {Pino}}]{jauma2024exploring}%
  \BibitemOpen
  \bibfield  {author} {\bibinfo {author} {\bibfnamefont {G.}~\bibnamefont
  {Jaum{\`a}}}, \bibinfo {author} {\bibfnamefont {J.~J.}\ \bibnamefont
  {Garc{\'\i}a-Ripoll}},\ and\ \bibinfo {author} {\bibfnamefont
  {M.}~\bibnamefont {Pino}},\ }\bibfield  {title} {\bibinfo {title} {Exploring
  quantum annealing architectures: A spin glass perspective},\ }\href@noop {}
  {\bibfield  {journal} {\bibinfo  {journal} {Advanced Quantum Technologies}\
  }\textbf {\bibinfo {volume} {7}},\ \bibinfo {pages} {2300245} (\bibinfo
  {year} {2024})}\BibitemShut {NoStop}%
\bibitem [{\citenamefont {Tiunov}\ \emph {et~al.}(2019)\citenamefont {Tiunov},
  \citenamefont {Ulanov},\ and\ \citenamefont {Lvovsky}}]{tiunov2019annealing}%
  \BibitemOpen
  \bibfield  {author} {\bibinfo {author} {\bibfnamefont {E.~S.}\ \bibnamefont
  {Tiunov}}, \bibinfo {author} {\bibfnamefont {A.~E.}\ \bibnamefont {Ulanov}},\
  and\ \bibinfo {author} {\bibfnamefont {A.}~\bibnamefont {Lvovsky}},\
  }\bibfield  {title} {\bibinfo {title} {Annealing by simulating the coherent
  {I}sing machine},\ }\href@noop {} {\bibfield  {journal} {\bibinfo  {journal}
  {Optics Express}\ }\textbf {\bibinfo {volume} {27}},\ \bibinfo {pages}
  {10288} (\bibinfo {year} {2019})}\BibitemShut {NoStop}%
\bibitem [{\citenamefont {Leleu}\ \emph {et~al.}(2021)\citenamefont {Leleu},
  \citenamefont {Khoyratee}, \citenamefont {Levi}, \citenamefont {Hamerly},
  \citenamefont {Kohno},\ and\ \citenamefont {Aihara}}]{leleu2021scaling}%
  \BibitemOpen
  \bibfield  {author} {\bibinfo {author} {\bibfnamefont {T.}~\bibnamefont
  {Leleu}}, \bibinfo {author} {\bibfnamefont {F.}~\bibnamefont {Khoyratee}},
  \bibinfo {author} {\bibfnamefont {T.}~\bibnamefont {Levi}}, \bibinfo {author}
  {\bibfnamefont {R.}~\bibnamefont {Hamerly}}, \bibinfo {author} {\bibfnamefont
  {T.}~\bibnamefont {Kohno}},\ and\ \bibinfo {author} {\bibfnamefont
  {K.}~\bibnamefont {Aihara}},\ }\bibfield  {title} {\bibinfo {title} {Scaling
  advantage of chaotic amplitude control for high-performance combinatorial
  optimization},\ }\href@noop {} {\bibfield  {journal} {\bibinfo  {journal}
  {Communications Physics}\ }\textbf {\bibinfo {volume} {4}},\ \bibinfo {pages}
  {266} (\bibinfo {year} {2021})}\BibitemShut {NoStop}%
\bibitem [{\citenamefont {Goto}\ \emph {et~al.}(2019)\citenamefont {Goto},
  \citenamefont {Tatsumura},\ and\ \citenamefont
  {Dixon}}]{goto2019combinatorial}%
  \BibitemOpen
  \bibfield  {author} {\bibinfo {author} {\bibfnamefont {H.}~\bibnamefont
  {Goto}}, \bibinfo {author} {\bibfnamefont {K.}~\bibnamefont {Tatsumura}},\
  and\ \bibinfo {author} {\bibfnamefont {A.~R.}\ \bibnamefont {Dixon}},\
  }\bibfield  {title} {\bibinfo {title} {Combinatorial optimization by
  simulating adiabatic bifurcations in nonlinear hamiltonian systems},\
  }\href@noop {} {\bibfield  {journal} {\bibinfo  {journal} {Science Advances}\
  }\textbf {\bibinfo {volume} {5}},\ \bibinfo {pages} {eaav2372} (\bibinfo
  {year} {2019})}\BibitemShut {NoStop}%
\bibitem [{\citenamefont {Goto}\ \emph {et~al.}(2021)\citenamefont {Goto},
  \citenamefont {Endo}, \citenamefont {Suzuki}, \citenamefont {Sakai},
  \citenamefont {Kanao}, \citenamefont {Hamakawa}, \citenamefont {Hidaka},
  \citenamefont {Yamasaki},\ and\ \citenamefont {Tatsumura}}]{goto2021high}%
  \BibitemOpen
  \bibfield  {author} {\bibinfo {author} {\bibfnamefont {H.}~\bibnamefont
  {Goto}}, \bibinfo {author} {\bibfnamefont {K.}~\bibnamefont {Endo}}, \bibinfo
  {author} {\bibfnamefont {M.}~\bibnamefont {Suzuki}}, \bibinfo {author}
  {\bibfnamefont {Y.}~\bibnamefont {Sakai}}, \bibinfo {author} {\bibfnamefont
  {T.}~\bibnamefont {Kanao}}, \bibinfo {author} {\bibfnamefont
  {Y.}~\bibnamefont {Hamakawa}}, \bibinfo {author} {\bibfnamefont
  {R.}~\bibnamefont {Hidaka}}, \bibinfo {author} {\bibfnamefont
  {M.}~\bibnamefont {Yamasaki}},\ and\ \bibinfo {author} {\bibfnamefont
  {K.}~\bibnamefont {Tatsumura}},\ }\bibfield  {title} {\bibinfo {title}
  {High-performance combinatorial optimization based on classical mechanics},\
  }\href@noop {} {\bibfield  {journal} {\bibinfo  {journal} {Science Advances}\
  }\textbf {\bibinfo {volume} {7}},\ \bibinfo {pages} {eabe7953} (\bibinfo
  {year} {2021})}\BibitemShut {NoStop}%
\bibitem [{Note2()}]{Note2}%
  \BibitemOpen
  \bibinfo {note} {It is often the case in industrial contexts that the optimal
  solution is not necessary; \protect \textit {reduced cost} solutions are
  often sufficient.}\BibitemShut {Stop}%
\bibitem [{\citenamefont {Or{\'u}s}(2014)}]{orus2014practical}%
  \BibitemOpen
  \bibfield  {author} {\bibinfo {author} {\bibfnamefont {R.}~\bibnamefont
  {Or{\'u}s}},\ }\bibfield  {title} {\bibinfo {title} {A practical introduction
  to tensor networks: Matrix product states and projected entangled pair
  states},\ }\href@noop {} {\bibfield  {journal} {\bibinfo  {journal} {Annals
  of Physics}\ }\textbf {\bibinfo {volume} {349}},\ \bibinfo {pages} {117}
  (\bibinfo {year} {2014})}\BibitemShut {NoStop}%
\bibitem [{\citenamefont {Ba{\~n}uls}(2023)}]{banuls2023tensor}%
  \BibitemOpen
  \bibfield  {author} {\bibinfo {author} {\bibfnamefont {M.~C.}\ \bibnamefont
  {Ba{\~n}uls}},\ }\bibfield  {title} {\bibinfo {title} {Tensor network
  algorithms: A route map},\ }\href@noop {} {\bibfield  {journal} {\bibinfo
  {journal} {Annual Review of Condensed Matter Physics}\ }\textbf {\bibinfo
  {volume} {14}},\ \bibinfo {pages} {173} (\bibinfo {year} {2023})}\BibitemShut
  {NoStop}%
\bibitem [{\citenamefont {Nishino}(1995)}]{nishino1995density}%
  \BibitemOpen
  \bibfield  {author} {\bibinfo {author} {\bibfnamefont {T.}~\bibnamefont
  {Nishino}},\ }\bibfield  {title} {\bibinfo {title} {Density matrix
  renormalization group method for 2d classical models},\ }\href@noop {}
  {\bibfield  {journal} {\bibinfo  {journal} {Journal of the Physical Society
  of Japan}\ }\textbf {\bibinfo {volume} {64}},\ \bibinfo {pages} {3598}
  (\bibinfo {year} {1995})}\BibitemShut {NoStop}%
\bibitem [{\citenamefont {Nishino}\ and\ \citenamefont
  {Okunishi}(1995)}]{nishino1995product}%
  \BibitemOpen
  \bibfield  {author} {\bibinfo {author} {\bibfnamefont {T.}~\bibnamefont
  {Nishino}}\ and\ \bibinfo {author} {\bibfnamefont {K.}~\bibnamefont
  {Okunishi}},\ }\bibfield  {title} {\bibinfo {title} {Product wave function
  renormalization group},\ }\href@noop {} {\bibfield  {journal} {\bibinfo
  {journal} {Journal of the Physical Society of Japan}\ }\textbf {\bibinfo
  {volume} {64}},\ \bibinfo {pages} {4084} (\bibinfo {year}
  {1995})}\BibitemShut {NoStop}%
\bibitem [{\citenamefont {Nishino}\ and\ \citenamefont
  {Okunishi}(1998)}]{nishino1998density}%
  \BibitemOpen
  \bibfield  {author} {\bibinfo {author} {\bibfnamefont {T.}~\bibnamefont
  {Nishino}}\ and\ \bibinfo {author} {\bibfnamefont {K.}~\bibnamefont
  {Okunishi}},\ }\bibfield  {title} {\bibinfo {title} {A density matrix
  algorithm for 3d classical models},\ }\href@noop {} {\bibfield  {journal}
  {\bibinfo  {journal} {Journal of the Physical Society of Japan}\ }\textbf
  {\bibinfo {volume} {67}},\ \bibinfo {pages} {3066} (\bibinfo {year}
  {1998})}\BibitemShut {NoStop}%
\bibitem [{\citenamefont {Murg}\ \emph {et~al.}(2005)\citenamefont {Murg},
  \citenamefont {Verstraete},\ and\ \citenamefont {Cirac}}]{murg2005efficient}%
  \BibitemOpen
  \bibfield  {author} {\bibinfo {author} {\bibfnamefont {V.}~\bibnamefont
  {Murg}}, \bibinfo {author} {\bibfnamefont {F.}~\bibnamefont {Verstraete}},\
  and\ \bibinfo {author} {\bibfnamefont {J.~I.}\ \bibnamefont {Cirac}},\
  }\bibfield  {title} {\bibinfo {title} {Efficient evaluation of partition
  functions of inhomogeneous many-body spin systems},\ }\href@noop {}
  {\bibfield  {journal} {\bibinfo  {journal} {Physical Review Letters}\
  }\textbf {\bibinfo {volume} {95}},\ \bibinfo {pages} {057206} (\bibinfo
  {year} {2005})}\BibitemShut {NoStop}%
\bibitem [{\citenamefont {Verstraete}\ \emph {et~al.}(2008)\citenamefont
  {Verstraete}, \citenamefont {Murg},\ and\ \citenamefont
  {Cirac}}]{verstraete2008matrix}%
  \BibitemOpen
  \bibfield  {author} {\bibinfo {author} {\bibfnamefont {F.}~\bibnamefont
  {Verstraete}}, \bibinfo {author} {\bibfnamefont {V.}~\bibnamefont {Murg}},\
  and\ \bibinfo {author} {\bibfnamefont {J.~I.}\ \bibnamefont {Cirac}},\
  }\bibfield  {title} {\bibinfo {title} {Matrix product states, projected
  entangled pair states, and variational renormalization group methods for
  quantum spin systems},\ }\href@noop {} {\bibfield  {journal} {\bibinfo
  {journal} {Advances in physics}\ }\textbf {\bibinfo {volume} {57}},\ \bibinfo
  {pages} {143} (\bibinfo {year} {2008})}\BibitemShut {NoStop}%
\bibitem [{\citenamefont {Rams}\ \emph {et~al.}(2021)\citenamefont {Rams},
  \citenamefont {Mohseni}, \citenamefont {Eppens}, \citenamefont
  {Ja{\l}owiecki},\ and\ \citenamefont {Gardas}}]{rams2021approximate}%
  \BibitemOpen
  \bibfield  {author} {\bibinfo {author} {\bibfnamefont {M.~M.}\ \bibnamefont
  {Rams}}, \bibinfo {author} {\bibfnamefont {M.}~\bibnamefont {Mohseni}},
  \bibinfo {author} {\bibfnamefont {D.}~\bibnamefont {Eppens}}, \bibinfo
  {author} {\bibfnamefont {K.}~\bibnamefont {Ja{\l}owiecki}},\ and\ \bibinfo
  {author} {\bibfnamefont {B.}~\bibnamefont {Gardas}},\ }\bibfield  {title}
  {\bibinfo {title} {Approximate optimization, sampling, and spin-glass droplet
  discovery with tensor networks},\ }\href@noop {} {\bibfield  {journal}
  {\bibinfo  {journal} {Physical Review E}\ }\textbf {\bibinfo {volume}
  {104}},\ \bibinfo {pages} {025308} (\bibinfo {year} {2021})}\BibitemShut
  {NoStop}%
\bibitem [{\citenamefont {Ja{\l}owiecki}\ \emph {et~al.}(2021)\citenamefont
  {Ja{\l}owiecki}, \citenamefont {Rams},\ and\ \citenamefont
  {Gardas}}]{jalowiecki2021brute}%
  \BibitemOpen
  \bibfield  {author} {\bibinfo {author} {\bibfnamefont {K.}~\bibnamefont
  {Ja{\l}owiecki}}, \bibinfo {author} {\bibfnamefont {M.~M.}\ \bibnamefont
  {Rams}},\ and\ \bibinfo {author} {\bibfnamefont {B.}~\bibnamefont {Gardas}},\
  }\bibfield  {title} {\bibinfo {title} {Brute-forcing spin-glass problems with
  cuda},\ }\href@noop {} {\bibfield  {journal} {\bibinfo  {journal} {Computer
  Physics Communications}\ }\textbf {\bibinfo {volume} {260}},\ \bibinfo
  {pages} {107728} (\bibinfo {year} {2021})}\BibitemShut {NoStop}%
\bibitem [{\citenamefont {Liu}\ \emph {et~al.}(2021)\citenamefont {Liu},
  \citenamefont {Wang},\ and\ \citenamefont {Zhang}}]{liu2021tropical}%
  \BibitemOpen
  \bibfield  {author} {\bibinfo {author} {\bibfnamefont {J.-G.}\ \bibnamefont
  {Liu}}, \bibinfo {author} {\bibfnamefont {L.}~\bibnamefont {Wang}},\ and\
  \bibinfo {author} {\bibfnamefont {P.}~\bibnamefont {Zhang}},\ }\bibfield
  {title} {\bibinfo {title} {Tropical tensor network for ground states of spin
  glasses},\ }\href@noop {} {\bibfield  {journal} {\bibinfo  {journal}
  {Physical Review Letters}\ }\textbf {\bibinfo {volume} {126}},\ \bibinfo
  {pages} {090506} (\bibinfo {year} {2021})}\BibitemShut {NoStop}%
\bibitem [{\citenamefont {Lanthier}\ \emph {et~al.}(2024)\citenamefont
  {Lanthier}, \citenamefont {C{\^o}t{\'e}},\ and\ \citenamefont
  {Kourtis}}]{lanthier2024tensor}%
  \BibitemOpen
  \bibfield  {author} {\bibinfo {author} {\bibfnamefont {B.}~\bibnamefont
  {Lanthier}}, \bibinfo {author} {\bibfnamefont {J.}~\bibnamefont
  {C{\^o}t{\'e}}},\ and\ \bibinfo {author} {\bibfnamefont {S.}~\bibnamefont
  {Kourtis}},\ }\bibfield  {title} {\bibinfo {title} {Tensor networks for $ p
  $-spin models},\ }\href@noop {} {\bibfield  {journal} {\bibinfo  {journal}
  {arXiv preprint arXiv:2405.08106}\ } (\bibinfo {year} {2024})}\BibitemShut
  {NoStop}%
\bibitem [{\citenamefont {Pan}\ \emph {et~al.}(2020)\citenamefont {Pan},
  \citenamefont {Zhou}, \citenamefont {Li},\ and\ \citenamefont
  {Zhang}}]{pan2020contracting}%
  \BibitemOpen
  \bibfield  {author} {\bibinfo {author} {\bibfnamefont {F.}~\bibnamefont
  {Pan}}, \bibinfo {author} {\bibfnamefont {P.}~\bibnamefont {Zhou}}, \bibinfo
  {author} {\bibfnamefont {S.}~\bibnamefont {Li}},\ and\ \bibinfo {author}
  {\bibfnamefont {P.}~\bibnamefont {Zhang}},\ }\bibfield  {title} {\bibinfo
  {title} {Contracting arbitrary tensor networks: general approximate algorithm
  and applications in graphical models and quantum circuit simulations},\
  }\href@noop {} {\bibfield  {journal} {\bibinfo  {journal} {Physical Review
  Letters}\ }\textbf {\bibinfo {volume} {125}},\ \bibinfo {pages} {060503}
  (\bibinfo {year} {2020})}\BibitemShut {NoStop}%
\bibitem [{\citenamefont {Gray}\ and\ \citenamefont
  {Chan}(2024)}]{gray2024hyperoptimized}%
  \BibitemOpen
  \bibfield  {author} {\bibinfo {author} {\bibfnamefont {J.}~\bibnamefont
  {Gray}}\ and\ \bibinfo {author} {\bibfnamefont {G.~K.-L.}\ \bibnamefont
  {Chan}},\ }\bibfield  {title} {\bibinfo {title} {Hyperoptimized approximate
  contraction of tensor networks with arbitrary geometry},\ }\href@noop {}
  {\bibfield  {journal} {\bibinfo  {journal} {Physical Review X}\ }\textbf
  {\bibinfo {volume} {14}},\ \bibinfo {pages} {011009} (\bibinfo {year}
  {2024})}\BibitemShut {NoStop}%
\bibitem [{\citenamefont {Ma}\ \emph {et~al.}(2024)\citenamefont {Ma},
  \citenamefont {Fishman}, \citenamefont {Stoudenmire},\ and\ \citenamefont
  {Solomonik}}]{ma2024approximate}%
  \BibitemOpen
  \bibfield  {author} {\bibinfo {author} {\bibfnamefont {L.}~\bibnamefont
  {Ma}}, \bibinfo {author} {\bibfnamefont {M.~T.}\ \bibnamefont {Fishman}},
  \bibinfo {author} {\bibfnamefont {E.~M.}\ \bibnamefont {Stoudenmire}},\ and\
  \bibinfo {author} {\bibfnamefont {E.}~\bibnamefont {Solomonik}},\ }\bibfield
  {title} {\bibinfo {title} {Approximate contraction of arbitrary tensor
  networks with a flexible and efficient density matrix algorithm},\
  }\href@noop {} {\bibfield  {journal} {\bibinfo  {journal} {arXiv preprint
  arXiv:2406.09769}\ } (\bibinfo {year} {2024})}\BibitemShut {NoStop}%
\bibitem [{Note3()}]{Note3}%
  \BibitemOpen
  \bibinfo {note} {Ref. \cite {gray2024hyperoptimized} shows that even in cases
  where an MPS-based contraction, such as the one used in Ref. \cite
  {rams2021approximate}, may be applied, the method of Ref. \cite
  {gray2024hyperoptimized} can significantly outperform MPS-based contraction
  in terms of both memory and time costs.}\BibitemShut {Stop}%
\bibitem [{Note4()}]{Note4}%
  \BibitemOpen
  \bibinfo {note} {The (assumed) distinction between P and NP does not preclude
  the existence of \protect \textit {heuristic} polynomial-time numerical
  methods for rugged-energy-landscape spin glasses---methods that have no
  performance guarantee regarding solution quality but in practice yield good
  solutions at least in some instances \cite {mohseni2022ising}}\BibitemShut
  {NoStop}%
\bibitem [{\citenamefont {Gray}(2018)}]{gray2018quimb}%
  \BibitemOpen
  \bibfield  {author} {\bibinfo {author} {\bibfnamefont {J.}~\bibnamefont
  {Gray}},\ }\bibfield  {title} {\bibinfo {title} {quimb: A python package for
  quantum information and many-body calculations},\ }\href@noop {} {\bibfield
  {journal} {\bibinfo  {journal} {Journal of Open Source Software}\ }\textbf
  {\bibinfo {volume} {3}},\ \bibinfo {pages} {819} (\bibinfo {year}
  {2018})}\BibitemShut {NoStop}%
\bibitem [{\citenamefont {Gray}\ and\ \citenamefont
  {Kourtis}(2021)}]{gray2021hyper}%
  \BibitemOpen
  \bibfield  {author} {\bibinfo {author} {\bibfnamefont {J.}~\bibnamefont
  {Gray}}\ and\ \bibinfo {author} {\bibfnamefont {S.}~\bibnamefont {Kourtis}},\
  }\bibfield  {title} {\bibinfo {title} {Hyper-optimized tensor network
  contraction},\ }\href@noop {} {\bibfield  {journal} {\bibinfo  {journal}
  {Quantum}\ }\textbf {\bibinfo {volume} {5}},\ \bibinfo {pages} {410}
  (\bibinfo {year} {2021})}\BibitemShut {NoStop}%
\bibitem [{\citenamefont {Perera}\ \emph
  {et~al.}(2020{\natexlab{a}})\citenamefont {Perera}, \citenamefont {Akpabio},
  \citenamefont {Hamze}, \citenamefont {Mandra}, \citenamefont {Rose},
  \citenamefont {Aramon},\ and\ \citenamefont {Katzgraber}}]{perera2020chook}%
  \BibitemOpen
  \bibfield  {author} {\bibinfo {author} {\bibfnamefont {D.}~\bibnamefont
  {Perera}}, \bibinfo {author} {\bibfnamefont {I.}~\bibnamefont {Akpabio}},
  \bibinfo {author} {\bibfnamefont {F.}~\bibnamefont {Hamze}}, \bibinfo
  {author} {\bibfnamefont {S.}~\bibnamefont {Mandra}}, \bibinfo {author}
  {\bibfnamefont {N.}~\bibnamefont {Rose}}, \bibinfo {author} {\bibfnamefont
  {M.}~\bibnamefont {Aramon}},\ and\ \bibinfo {author} {\bibfnamefont {H.~G.}\
  \bibnamefont {Katzgraber}},\ }\bibfield  {title} {\bibinfo {title} {Chook--a
  comprehensive suite for generating binary optimization problems with planted
  solutions},\ }\href@noop {} {\bibfield  {journal} {\bibinfo  {journal} {arXiv
  preprint arXiv:2005.14344}\ } (\bibinfo {year}
  {2020}{\natexlab{a}})}\BibitemShut {NoStop}%
\bibitem [{\citenamefont {Hamze}\ \emph {et~al.}(2018)\citenamefont {Hamze},
  \citenamefont {Jacob}, \citenamefont {Ochoa}, \citenamefont {Perera},
  \citenamefont {Wang},\ and\ \citenamefont {Katzgraber}}]{hamze2018near}%
  \BibitemOpen
  \bibfield  {author} {\bibinfo {author} {\bibfnamefont {F.}~\bibnamefont
  {Hamze}}, \bibinfo {author} {\bibfnamefont {D.~C.}\ \bibnamefont {Jacob}},
  \bibinfo {author} {\bibfnamefont {A.~J.}\ \bibnamefont {Ochoa}}, \bibinfo
  {author} {\bibfnamefont {D.}~\bibnamefont {Perera}}, \bibinfo {author}
  {\bibfnamefont {W.}~\bibnamefont {Wang}},\ and\ \bibinfo {author}
  {\bibfnamefont {H.~G.}\ \bibnamefont {Katzgraber}},\ }\bibfield  {title}
  {\bibinfo {title} {From near to eternity: spin-glass planting, tiling
  puzzles, and constraint-satisfaction problems},\ }\href@noop {} {\bibfield
  {journal} {\bibinfo  {journal} {Physical Review E}\ }\textbf {\bibinfo
  {volume} {97}},\ \bibinfo {pages} {043303} (\bibinfo {year}
  {2018})}\BibitemShut {NoStop}%
\bibitem [{\citenamefont {Perera}\ \emph
  {et~al.}(2020{\natexlab{b}})\citenamefont {Perera}, \citenamefont {Hamze},
  \citenamefont {Raymond}, \citenamefont {Weigel},\ and\ \citenamefont
  {Katzgraber}}]{perera2020computational}%
  \BibitemOpen
  \bibfield  {author} {\bibinfo {author} {\bibfnamefont {D.}~\bibnamefont
  {Perera}}, \bibinfo {author} {\bibfnamefont {F.}~\bibnamefont {Hamze}},
  \bibinfo {author} {\bibfnamefont {J.}~\bibnamefont {Raymond}}, \bibinfo
  {author} {\bibfnamefont {M.}~\bibnamefont {Weigel}},\ and\ \bibinfo {author}
  {\bibfnamefont {H.~G.}\ \bibnamefont {Katzgraber}},\ }\bibfield  {title}
  {\bibinfo {title} {Computational hardness of spin-glass problems with
  tile-planted solutions},\ }\href@noop {} {\bibfield  {journal} {\bibinfo
  {journal} {Physical Review E}\ }\textbf {\bibinfo {volume} {101}},\ \bibinfo
  {pages} {023316} (\bibinfo {year} {2020}{\natexlab{b}})}\BibitemShut
  {NoStop}%
\bibitem [{\citenamefont {Fan}\ \emph {et~al.}(2023)\citenamefont {Fan},
  \citenamefont {Shen}, \citenamefont {Nussinov}, \citenamefont {Liu},
  \citenamefont {Sun},\ and\ \citenamefont {Liu}}]{fan2023searching}%
  \BibitemOpen
  \bibfield  {author} {\bibinfo {author} {\bibfnamefont {C.}~\bibnamefont
  {Fan}}, \bibinfo {author} {\bibfnamefont {M.}~\bibnamefont {Shen}}, \bibinfo
  {author} {\bibfnamefont {Z.}~\bibnamefont {Nussinov}}, \bibinfo {author}
  {\bibfnamefont {Z.}~\bibnamefont {Liu}}, \bibinfo {author} {\bibfnamefont
  {Y.}~\bibnamefont {Sun}},\ and\ \bibinfo {author} {\bibfnamefont {Y.-Y.}\
  \bibnamefont {Liu}},\ }\bibfield  {title} {\bibinfo {title} {Searching for
  spin glass ground states through deep reinforcement learning},\ }\href@noop
  {} {\bibfield  {journal} {\bibinfo  {journal} {Nature communications}\
  }\textbf {\bibinfo {volume} {14}},\ \bibinfo {pages} {725} (\bibinfo {year}
  {2023})}\BibitemShut {NoStop}%
\bibitem [{\citenamefont {Gangat}()}]{AAG}%
  \BibitemOpen
  \bibfield  {author} {\bibinfo {author} {\bibfnamefont {A.~A.}\ \bibnamefont
  {Gangat}},\ }\href@noop {} {\bibinfo {title} {(in preparation)}}\BibitemShut
  {NoStop}%
\bibitem [{\citenamefont {Susa}\ \emph {et~al.}(2018)\citenamefont {Susa},
  \citenamefont {Yamashiro}, \citenamefont {Yamamoto},\ and\ \citenamefont
  {Nishimori}}]{susa2018exponential}%
  \BibitemOpen
  \bibfield  {author} {\bibinfo {author} {\bibfnamefont {Y.}~\bibnamefont
  {Susa}}, \bibinfo {author} {\bibfnamefont {Y.}~\bibnamefont {Yamashiro}},
  \bibinfo {author} {\bibfnamefont {M.}~\bibnamefont {Yamamoto}},\ and\
  \bibinfo {author} {\bibfnamefont {H.}~\bibnamefont {Nishimori}},\ }\bibfield
  {title} {\bibinfo {title} {Exponential speedup of quantum annealing by
  inhomogeneous driving of the transverse field},\ }\href@noop {} {\bibfield
  {journal} {\bibinfo  {journal} {Journal of the Physical Society of Japan}\
  }\textbf {\bibinfo {volume} {87}},\ \bibinfo {pages} {023002} (\bibinfo
  {year} {2018})}\BibitemShut {NoStop}%
\bibitem [{\citenamefont {Raymond}\ \emph {et~al.}(2023)\citenamefont
  {Raymond}, \citenamefont {Stevanovic}, \citenamefont {Bernoudy},
  \citenamefont {Boothby}, \citenamefont {McGeoch}, \citenamefont {Berkley},
  \citenamefont {Farr{\'e}}, \citenamefont {Pasvolsky},\ and\ \citenamefont
  {King}}]{raymond2023hybrid}%
  \BibitemOpen
  \bibfield  {author} {\bibinfo {author} {\bibfnamefont {J.}~\bibnamefont
  {Raymond}}, \bibinfo {author} {\bibfnamefont {R.}~\bibnamefont {Stevanovic}},
  \bibinfo {author} {\bibfnamefont {W.}~\bibnamefont {Bernoudy}}, \bibinfo
  {author} {\bibfnamefont {K.}~\bibnamefont {Boothby}}, \bibinfo {author}
  {\bibfnamefont {C.~C.}\ \bibnamefont {McGeoch}}, \bibinfo {author}
  {\bibfnamefont {A.~J.}\ \bibnamefont {Berkley}}, \bibinfo {author}
  {\bibfnamefont {P.}~\bibnamefont {Farr{\'e}}}, \bibinfo {author}
  {\bibfnamefont {J.}~\bibnamefont {Pasvolsky}},\ and\ \bibinfo {author}
  {\bibfnamefont {A.~D.}\ \bibnamefont {King}},\ }\bibfield  {title} {\bibinfo
  {title} {Hybrid quantum annealing for larger-than-qpu lattice-structured
  problems},\ }\href@noop {} {\bibfield  {journal} {\bibinfo  {journal} {ACM
  Transactions on Quantum Computing}\ }\textbf {\bibinfo {volume} {4}},\
  \bibinfo {pages} {1} (\bibinfo {year} {2023})}\BibitemShut {NoStop}%
\bibitem [{\citenamefont {Katzgraber}\ \emph {et~al.}(2015)\citenamefont
  {Katzgraber}, \citenamefont {Hamze}, \citenamefont {Zhu}, \citenamefont
  {Ochoa},\ and\ \citenamefont {Munoz-Bauza}}]{katzgraber2015seeking}%
  \BibitemOpen
  \bibfield  {author} {\bibinfo {author} {\bibfnamefont {H.~G.}\ \bibnamefont
  {Katzgraber}}, \bibinfo {author} {\bibfnamefont {F.}~\bibnamefont {Hamze}},
  \bibinfo {author} {\bibfnamefont {Z.}~\bibnamefont {Zhu}}, \bibinfo {author}
  {\bibfnamefont {A.~J.}\ \bibnamefont {Ochoa}},\ and\ \bibinfo {author}
  {\bibfnamefont {H.}~\bibnamefont {Munoz-Bauza}},\ }\bibfield  {title}
  {\bibinfo {title} {Seeking quantum speedup through spin glasses: The good,
  the bad, and the ugly},\ }\href@noop {} {\bibfield  {journal} {\bibinfo
  {journal} {Physical Review X}\ }\textbf {\bibinfo {volume} {5}},\ \bibinfo
  {pages} {031026} (\bibinfo {year} {2015})}\BibitemShut {NoStop}%
\bibitem [{\citenamefont {De~las Cuevas}\ and\ \citenamefont
  {Cubitt}(2016)}]{de2016simple}%
  \BibitemOpen
  \bibfield  {author} {\bibinfo {author} {\bibfnamefont {G.}~\bibnamefont
  {De~las Cuevas}}\ and\ \bibinfo {author} {\bibfnamefont {T.~S.}\ \bibnamefont
  {Cubitt}},\ }\bibfield  {title} {\bibinfo {title} {Simple universal models
  capture all classical spin physics},\ }\href@noop {} {\bibfield  {journal}
  {\bibinfo  {journal} {Science}\ }\textbf {\bibinfo {volume} {351}},\ \bibinfo
  {pages} {1180} (\bibinfo {year} {2016})}\BibitemShut {NoStop}%
\bibitem [{\citenamefont {Zhang}(2023)}]{zhang2023mapping}%
  \BibitemOpen
  \bibfield  {author} {\bibinfo {author} {\bibfnamefont {Z.}~\bibnamefont
  {Zhang}},\ }\bibfield  {title} {\bibinfo {title} {Mapping between
  {S}pin-{G}lass {T}hree-{D}imensional (3{D}) {I}sing {M}odel and {B}oolean
  {S}atisfiability {P}roblem},\ }\href@noop {} {\bibfield  {journal} {\bibinfo
  {journal} {Mathematics}\ }\textbf {\bibinfo {volume} {11}},\ \bibinfo {pages}
  {237} (\bibinfo {year} {2023})}\BibitemShut {NoStop}%
\bibitem [{\citenamefont {Ferris}(2015)}]{ferris2015unbiased}%
  \BibitemOpen
  \bibfield  {author} {\bibinfo {author} {\bibfnamefont {A.~J.}\ \bibnamefont
  {Ferris}},\ }\bibfield  {title} {\bibinfo {title} {Unbiased monte carlo for
  the age of tensor networks},\ }\href@noop {} {\bibfield  {journal} {\bibinfo
  {journal} {arXiv preprint arXiv:1507.00767}\ } (\bibinfo {year}
  {2015})}\BibitemShut {NoStop}%
\bibitem [{\citenamefont {Huggins}\ \emph {et~al.}(2017)\citenamefont
  {Huggins}, \citenamefont {Freeman}, \citenamefont {Stoudenmire},
  \citenamefont {Tubman},\ and\ \citenamefont {Whaley}}]{huggins2017monte}%
  \BibitemOpen
  \bibfield  {author} {\bibinfo {author} {\bibfnamefont {W.}~\bibnamefont
  {Huggins}}, \bibinfo {author} {\bibfnamefont {C.~D.}\ \bibnamefont
  {Freeman}}, \bibinfo {author} {\bibfnamefont {M.}~\bibnamefont
  {Stoudenmire}}, \bibinfo {author} {\bibfnamefont {N.~M.}\ \bibnamefont
  {Tubman}},\ and\ \bibinfo {author} {\bibfnamefont {K.~B.}\ \bibnamefont
  {Whaley}},\ }\bibfield  {title} {\bibinfo {title} {Monte {C}arlo tensor
  network renormalization},\ }\href@noop {} {\bibfield  {journal} {\bibinfo
  {journal} {arXiv preprint arXiv:1710.03757}\ } (\bibinfo {year}
  {2017})}\BibitemShut {NoStop}%
\bibitem [{\citenamefont {Fr{\'\i}as~P{\'e}rez}\ \emph
  {et~al.}(2023)\citenamefont {Fr{\'\i}as~P{\'e}rez}, \citenamefont
  {Mari{\"e}n}, \citenamefont {P{\'e}rez~Garc{\'\i}a}, \citenamefont
  {Ba{\~n}uls},\ and\ \citenamefont {Iblisdir}}]{frias2023collective}%
  \BibitemOpen
  \bibfield  {author} {\bibinfo {author} {\bibfnamefont {M.}~\bibnamefont
  {Fr{\'\i}as~P{\'e}rez}}, \bibinfo {author} {\bibfnamefont {M.}~\bibnamefont
  {Mari{\"e}n}}, \bibinfo {author} {\bibfnamefont {D.}~\bibnamefont
  {P{\'e}rez~Garc{\'\i}a}}, \bibinfo {author} {\bibfnamefont {M.~C.}\
  \bibnamefont {Ba{\~n}uls}},\ and\ \bibinfo {author} {\bibfnamefont
  {S.}~\bibnamefont {Iblisdir}},\ }\bibfield  {title} {\bibinfo {title}
  {Collective {M}onte {C}arlo updates through tensor network renormalization},\
  }\href@noop {} {\bibfield  {journal} {\bibinfo  {journal} {SciPost Physics}\
  }\textbf {\bibinfo {volume} {14}},\ \bibinfo {pages} {123} (\bibinfo {year}
  {2023})}\BibitemShut {NoStop}%
\bibitem [{\citenamefont {Chen}\ \emph {et~al.}(2024)\citenamefont {Chen},
  \citenamefont {Guo}, \citenamefont {Zhang}, \citenamefont {Zhang},\ and\
  \citenamefont {Deng}}]{chen2024tensor}%
  \BibitemOpen
  \bibfield  {author} {\bibinfo {author} {\bibfnamefont {T.}~\bibnamefont
  {Chen}}, \bibinfo {author} {\bibfnamefont {E.}~\bibnamefont {Guo}}, \bibinfo
  {author} {\bibfnamefont {W.}~\bibnamefont {Zhang}}, \bibinfo {author}
  {\bibfnamefont {P.}~\bibnamefont {Zhang}},\ and\ \bibinfo {author}
  {\bibfnamefont {Y.}~\bibnamefont {Deng}},\ }\bibfield  {title} {\bibinfo
  {title} {Tensor network {M}onte {C}arlo simulations for the two-dimensional
  random-bond ising model},\ }\href@noop {} {\bibfield  {journal} {\bibinfo
  {journal} {arXiv preprint arXiv:2409.06538}\ } (\bibinfo {year}
  {2024})}\BibitemShut {NoStop}%
\bibitem [{\citenamefont {Charbonneau}\ \emph {et~al.}(2023)\citenamefont
  {Charbonneau}, \citenamefont {Marinari}, \citenamefont {M\'ezard},
  \citenamefont {Parisi}, \citenamefont {Ricci-{T}ersenghi}, \citenamefont
  {Sicuro},\ and\ \citenamefont {Zamponi}}]{charbonneau2023spin}%
  \BibitemOpen
  \bibfield  {author} {\bibinfo {author} {\bibfnamefont {P.}~\bibnamefont
  {Charbonneau}}, \bibinfo {author} {\bibfnamefont {E.}~\bibnamefont
  {Marinari}}, \bibinfo {author} {\bibfnamefont {M.}~\bibnamefont {M\'ezard}},
  \bibinfo {author} {\bibfnamefont {G.}~\bibnamefont {Parisi}}, \bibinfo
  {author} {\bibfnamefont {F.}~\bibnamefont {Ricci-{T}ersenghi}}, \bibinfo
  {author} {\bibfnamefont {G.}~\bibnamefont {Sicuro}},\ and\ \bibinfo {author}
  {\bibfnamefont {F.}~\bibnamefont {Zamponi}},\ }\href@noop {} {\emph {\bibinfo
  {title} {Spin Glass Theory and Far Beyond: Replica Symmetry Breaking after 40
  Years}}}\ (\bibinfo  {publisher} {World Scientific},\ \bibinfo {year}
  {2023})\BibitemShut {NoStop}%
\end{thebibliography}%

\appendix
\section{Tensor network example for conditional marginals}
\label{sec:TN_construction}
\begin{figure}[h]
\includegraphics[width=0.48\textwidth]{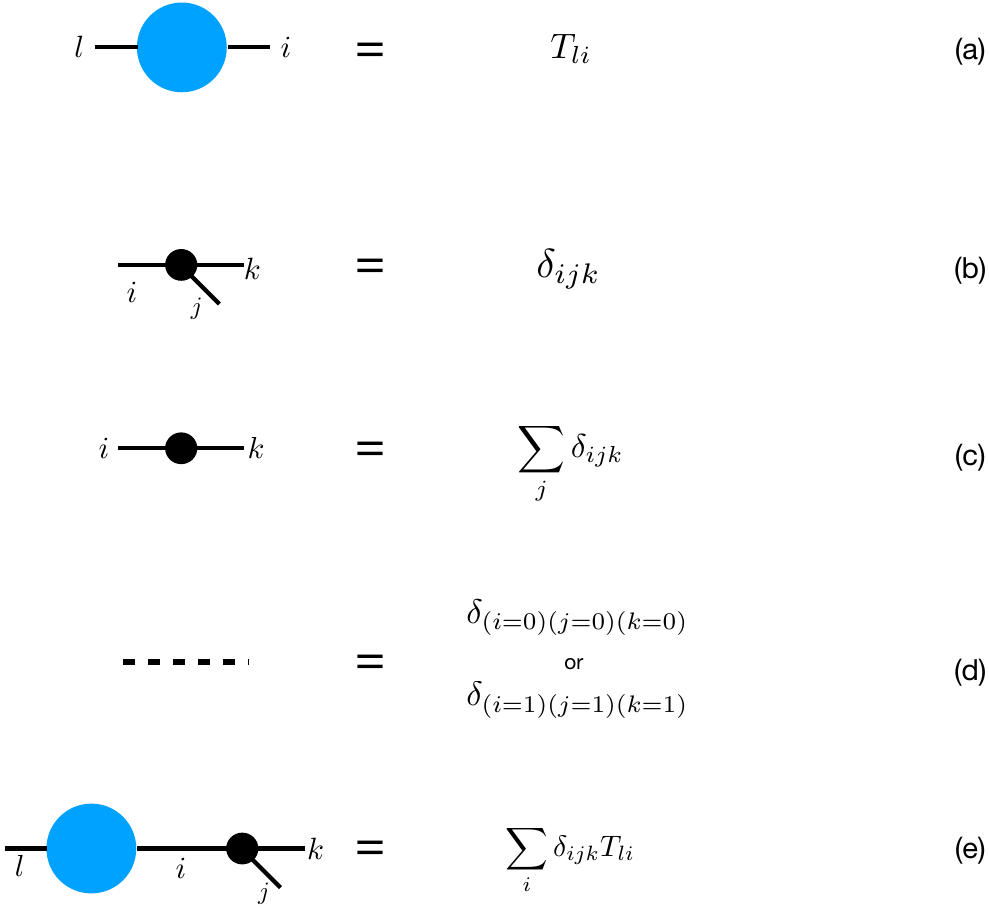}
\caption{(color online). Diagrammatic notation used for illustrating tensor network formulation and contraction for classical spin glasses.  In the case of Ising spins, all legs (i.e., indices) are of dimension two.}
\label{fig:TN_notation}
\end{figure}

We envision the square and cubic spin lattices as graphs with vertices at each spin.  The tensor network construction that we employ entails assigning a matrix to each edge of the graph.  The matrix encodes the possible Boltzmann weights for the joint configuration of the two spins that it connects.  For example, the matrix $T_{ij}$ assigned to the edge between spin $i$ and spin $j$ takes the form
\begin{equation}
T_{ij} = 
\begin{bmatrix}
e^{-\beta J_{ij}} & e^{\beta J_{ij}} \\
e^{\beta J_{ij}} & e^{-\beta J_{ij}} 
\end{bmatrix}.
\label{eqn:matrix}
\end{equation}
We represent these matrices with a larger (blue) circle and two legs (Fig. \ref{fig:TN_notation}a), where each leg corresponds to an index of the matrix.  This assignment of matrices on the square and cubic spin lattices results in a hyperindex at each spin, which is easily handled by the \texttt{quimb} \cite{gray2018quimb} and \texttt{cotengra} \cite{gray2021hyper} Python libraries that we use in this work.  However, for pedagogical clarity, in this section we replace each hyperindex with a kronecker delta function, which we represent with a smaller (black) circle and one leg attached to the smaller circle for each index (Fig. \ref{fig:TN_notation}b).  Deleting a leg (as in going from Fig. \ref{fig:TN_notation}b to Fig. \ref{fig:TN_notation}c) denotes summing over the corresponding index.  Deleting a smaller black circle and its dangling leg and replacing the remaining legs with dashed lines (as in going from Fig. \ref{fig:TN_notation}b to Fig. \ref{fig:TN_notation}d) represents a selection of a particular value of the corresponding index, which is equivalent to projecting the corresponding spin onto a particular state.  Joining legs together represents a contraction of the corresponding indices (Fig. \ref{fig:TN_notation}e). 

\begin{figure}[h]
\includegraphics[width=0.40\textwidth]{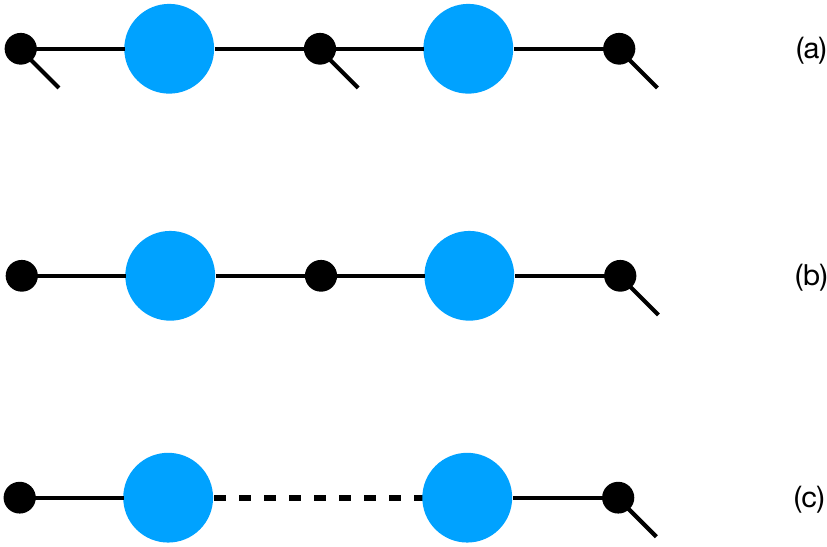}
\caption{(color online). Three-site classical Ising model example for marginal computations with tensor networks (see Fig. \ref{fig:TN_notation} for notation).  Small (black) circles are located at the sites of the spins. (a) Tensor network for marginal distribution of the entire lattice.  (b) Tensor network for the marginal distribution of the rightmost spin only.  See  second paragraph of Appendix \ref{sec:TN_construction} for further description.}
\label{fig:TN_example}
\end{figure}

To illustrate how this construction produces the desired marginals, we present an example for the three-site Ising chain with open boundaries.  In Fig. \ref{fig:TN_example}a, we show the tensor network whose contraction produces a single three-site tensor, which encodes the (unnormalized) marginal distribution for all three spins.  Selecting a particular value for a single open leg (i.e., uncontracted index) is equivalent to projecting the corresponding spin onto a particular state.  In the case of Ising spins, each leg (i.e., index) in the uncontracted tensor network has dimension two.  Selecting particular values for all of the dangling legs produces a tensor network whose contraction gives the (unnormalized) probability for the corresponding spin configuration of all three spins.  Summing over the left two dangling legs produces the tensor network in Fig. \ref{fig:TN_example}b, whose contraction yields a vector containing the (unnormalized) unconditional marginal distribution for the third spin.  Finally, summing over the first dangling leg and selecting a particular value for the middle dangling leg produces the tensor network in Fig. \ref{fig:TN_example}c, whose contraction yields the (unnormalized) \textit{conditional} marginal distribution for the third spin (conditioned upon the selected value for the second spin); the marginal comes out of the contraction as a two-element vector (i.e., a circle with one open leg).

To compute the most likely spin configuration for the full spin glass at a given value of $\beta$, the following procedure is used: 0) construct the tensor network corresponding to the marginal distribution for the entire spin lattice (call it TN0), 1) make a copy of TN0, and sum over all open indices except for the one corresponding to the first spin, 2) contract this modified copy of TN0 to produce the marginal for the first spin, 3) project the first spin in the original TN0 onto its most probable state as determined from the marginal produced in the previous step, 4) use this updated version of TN0 to perform steps (1)-(3) for the next spin.

\newpage
\section{Square-lattice solution quality}
\label{sec:TP2D_data}

\begin{figure*}
    \subfloat[$C_1, L=20$]{{\includegraphics[width=0.47\textwidth]{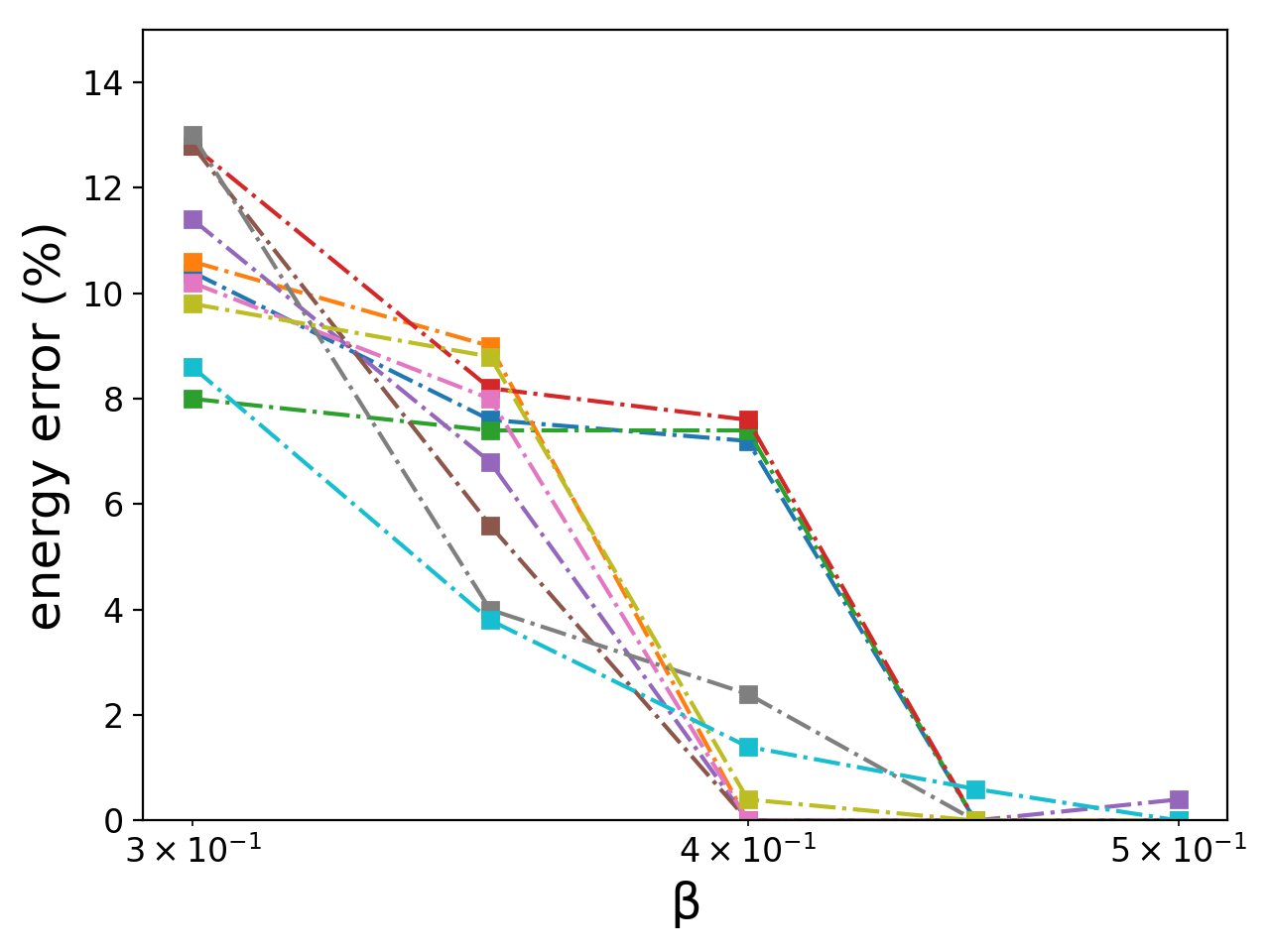} }} \hspace{0.4cm}
    \subfloat[$C_2, L=20$]{{\includegraphics[width=0.47\textwidth]{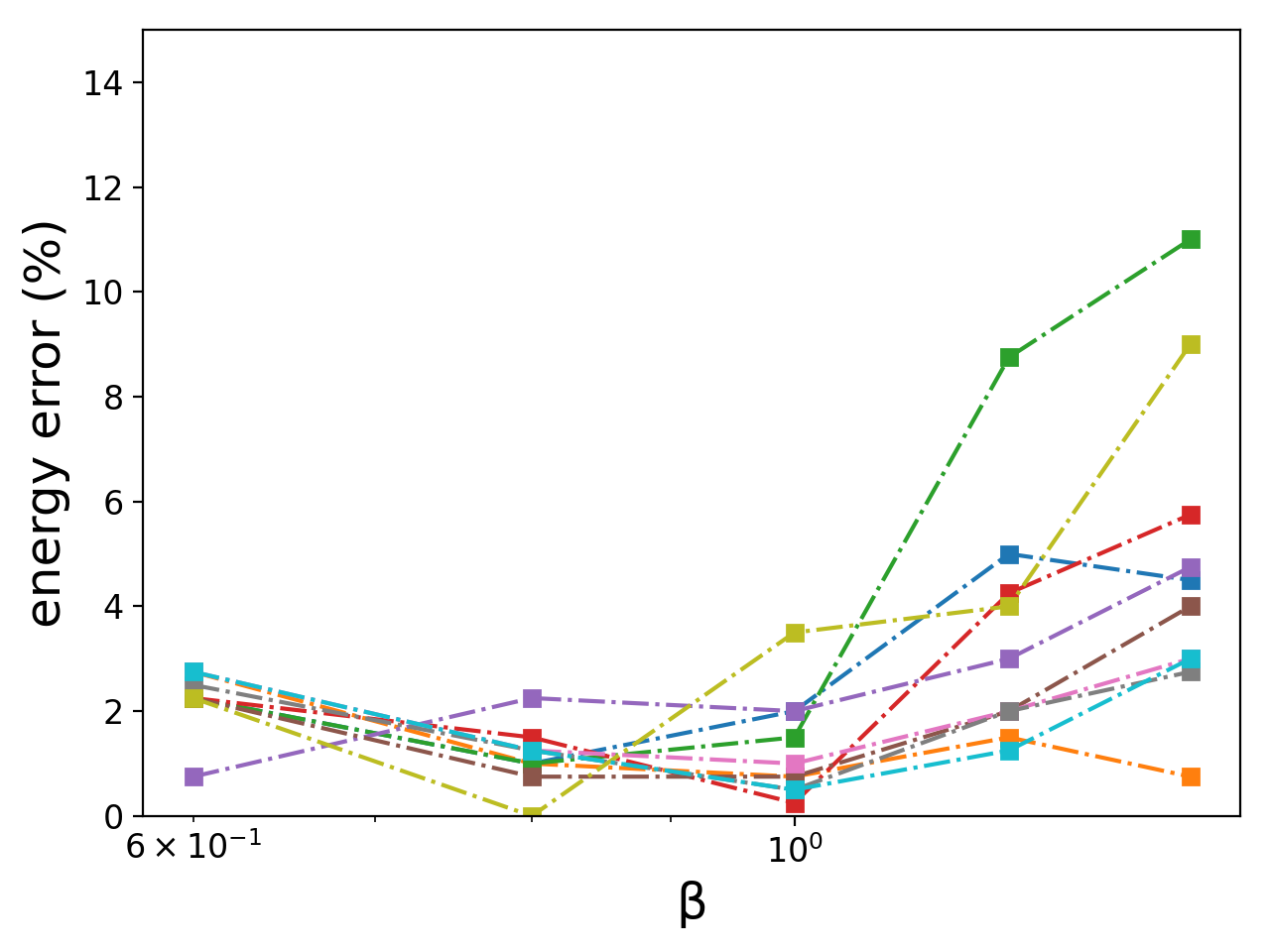} }}
    
    \subfloat[$C_1, L=32$]{{\includegraphics[width=0.47\textwidth]{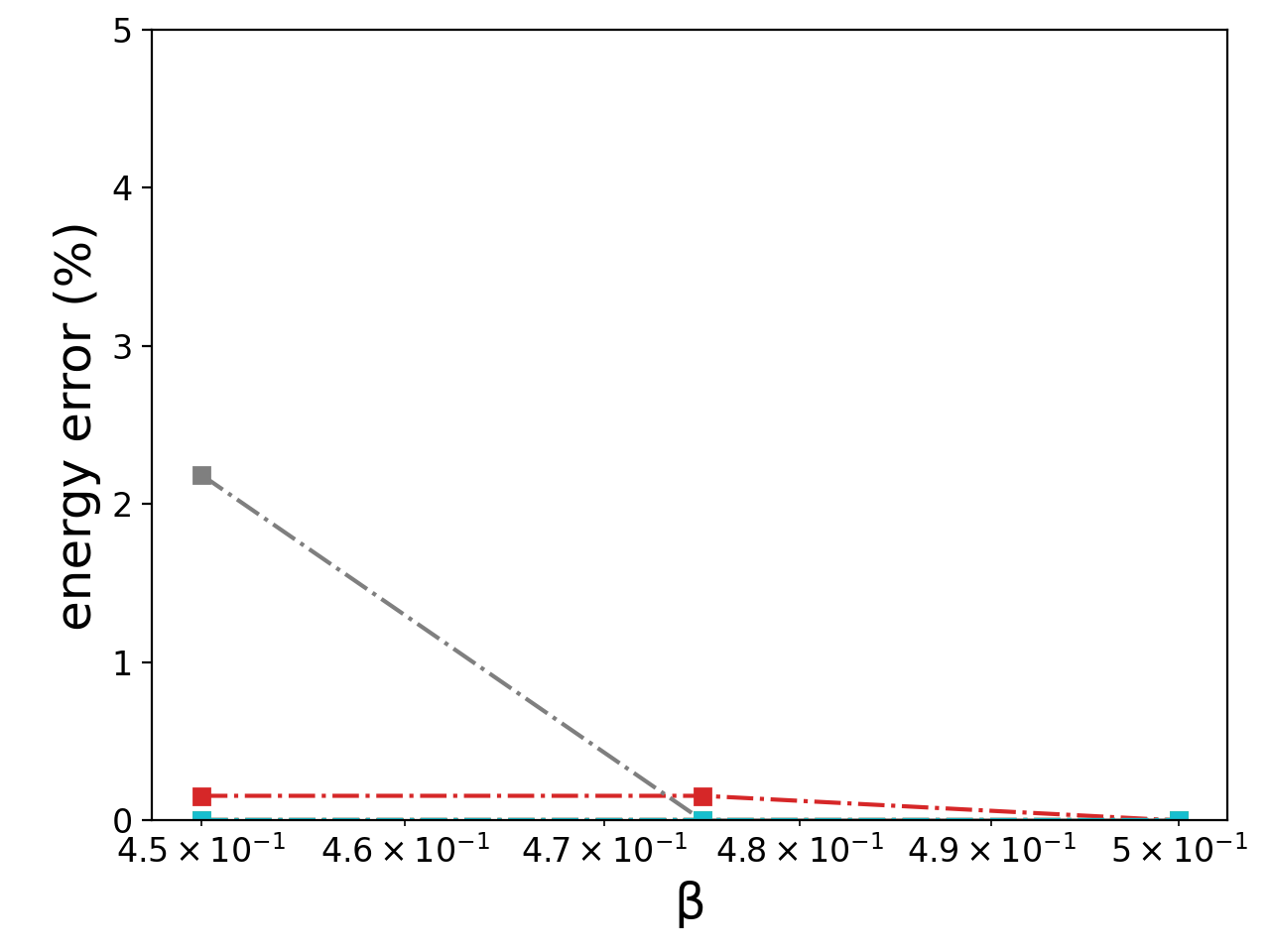} }} \hspace{0.4cm}
    \subfloat[$C_2, L=32$]{{\includegraphics[width=0.47\textwidth]{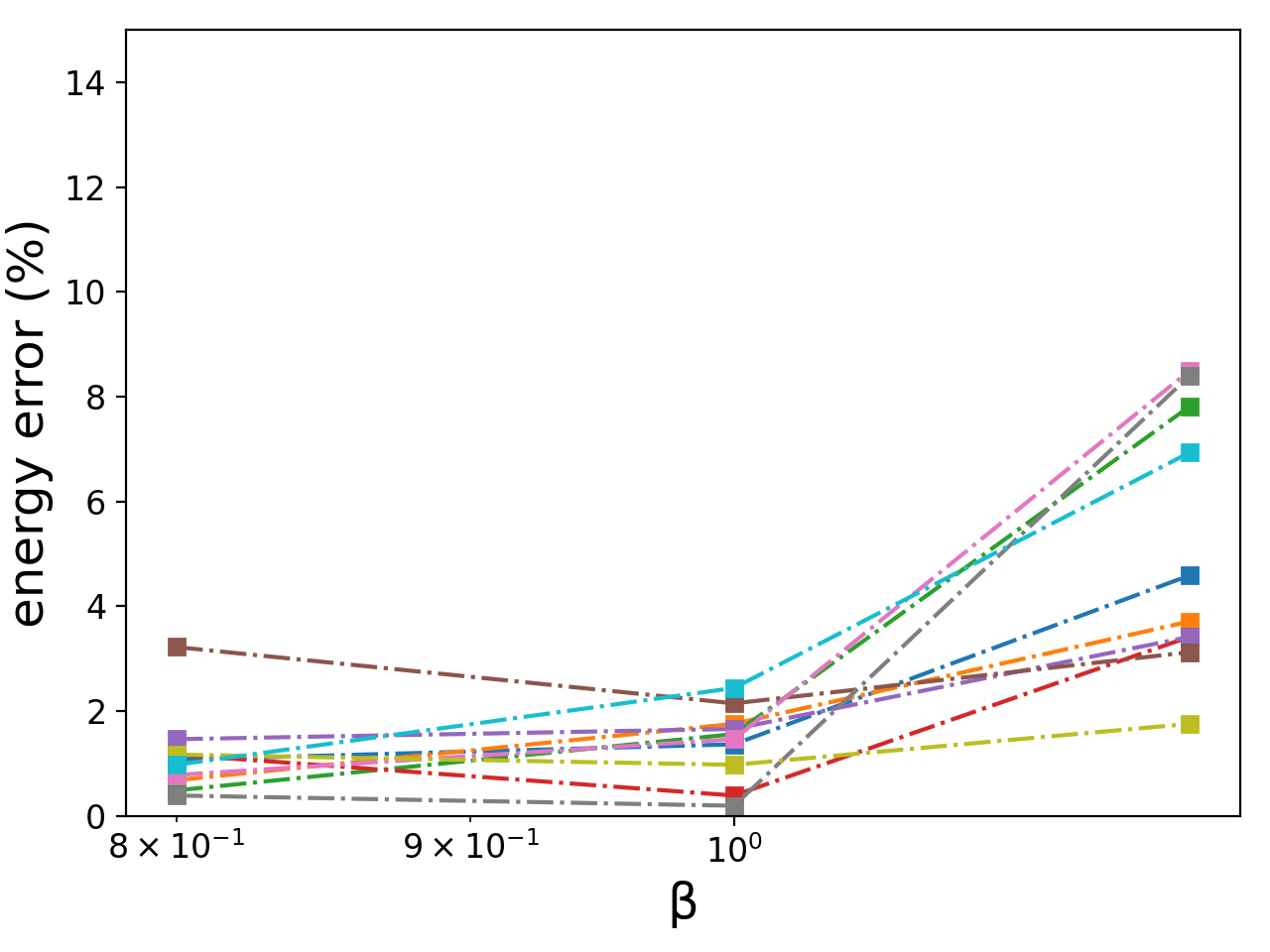} }}

    \subfloat[$C_1, L=48$]{{\includegraphics[width=0.47\textwidth]{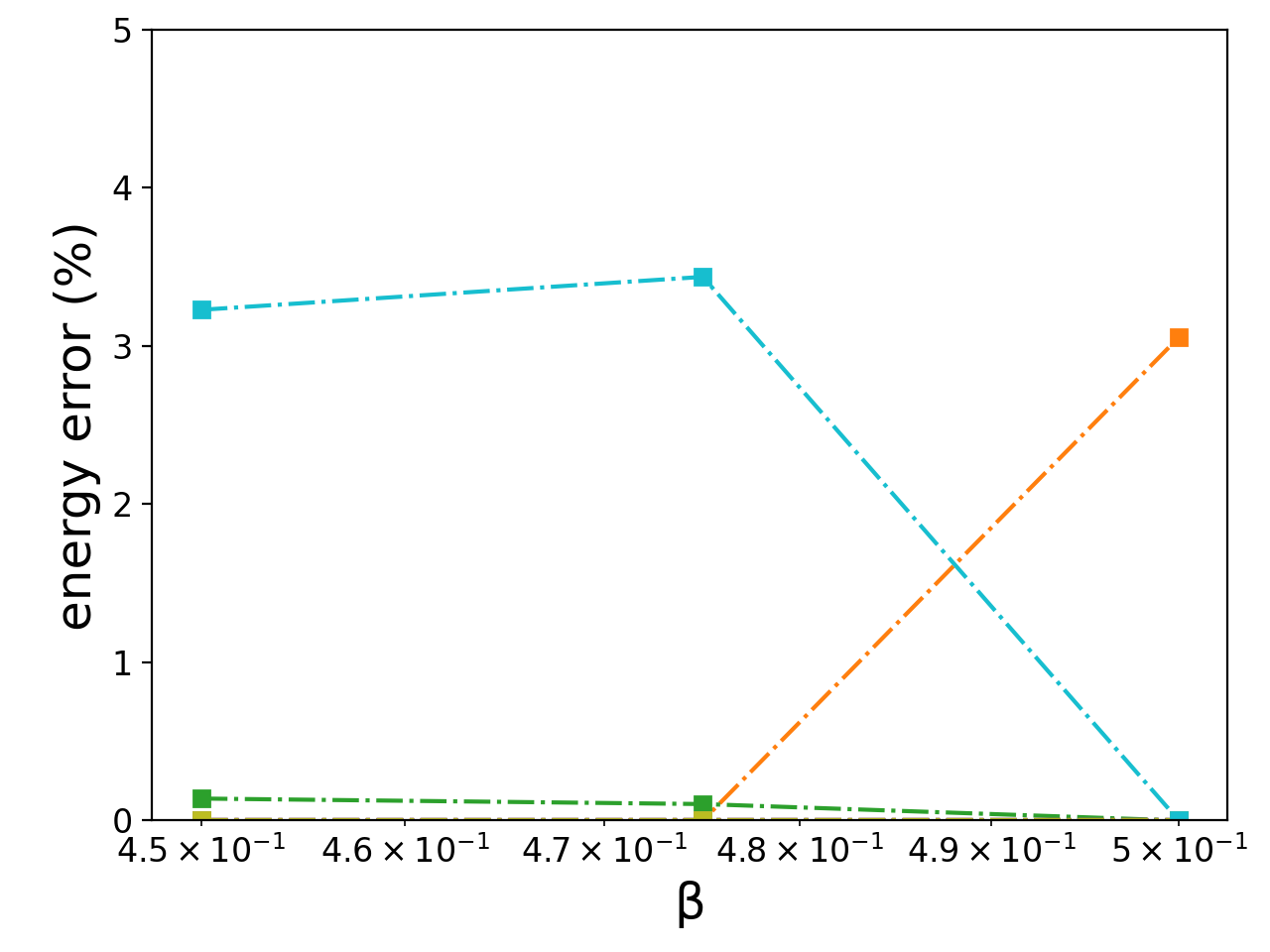} }} \hspace{0.4cm}
    \subfloat[$C_2, L=48$]{{\includegraphics[width=0.47\textwidth]{fig_error_C2_48x48.png} }}
    
    \caption{(color online). $C_1$ (left column) and $C_2$ (right column): solution quality vs. $\beta$, ten instances with $\chi=4$.  Data points belonging to the same instance are connected by dash-dotted lines.}
\end{figure*}

\begin{figure*}
    \subfloat[$C_3, L=20$]{{\includegraphics[width=0.47\textwidth]{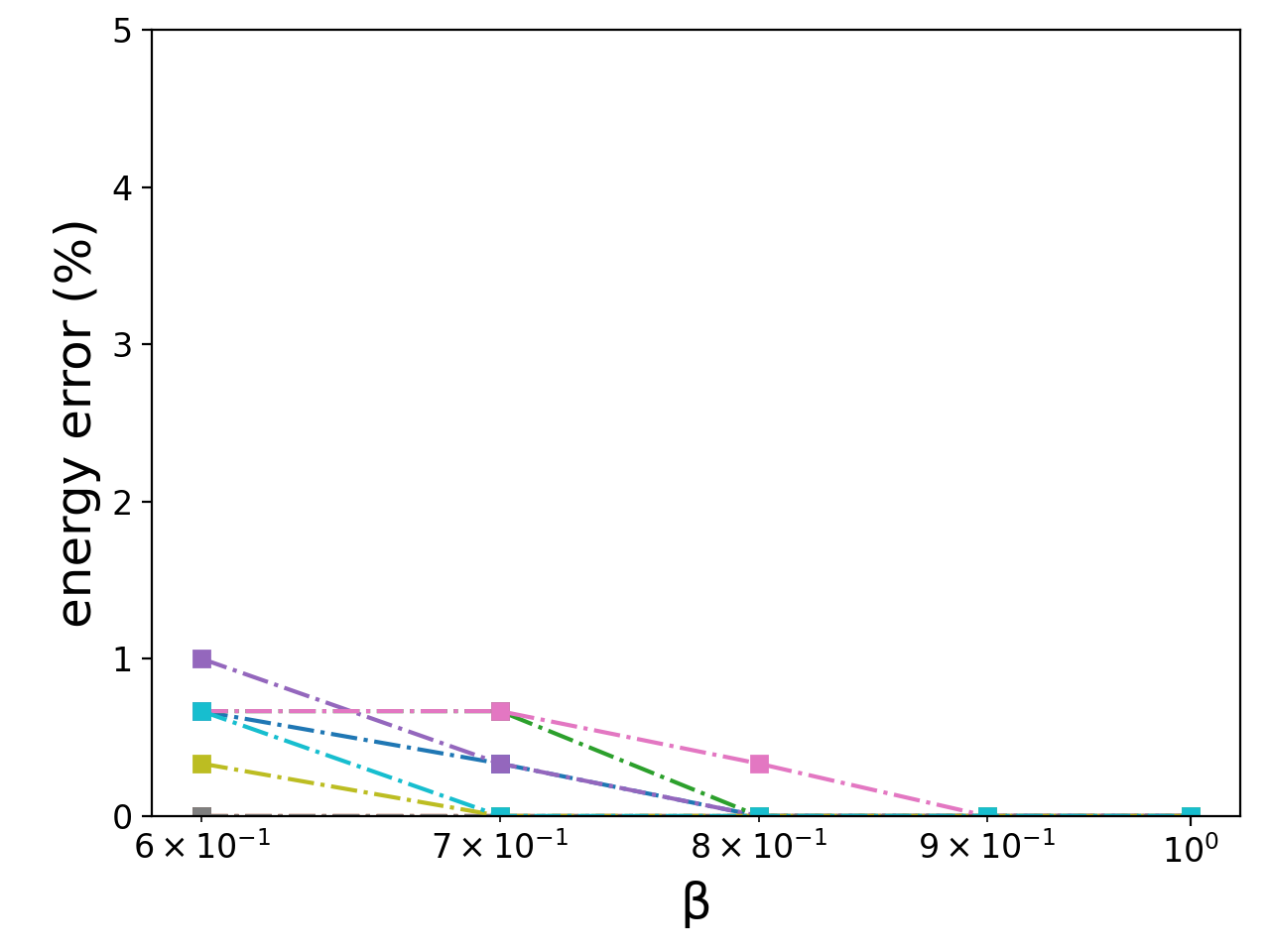} }} \hspace{0.4cm}
    \subfloat[$C_4, L=20$]{{\includegraphics[width=0.47\textwidth]{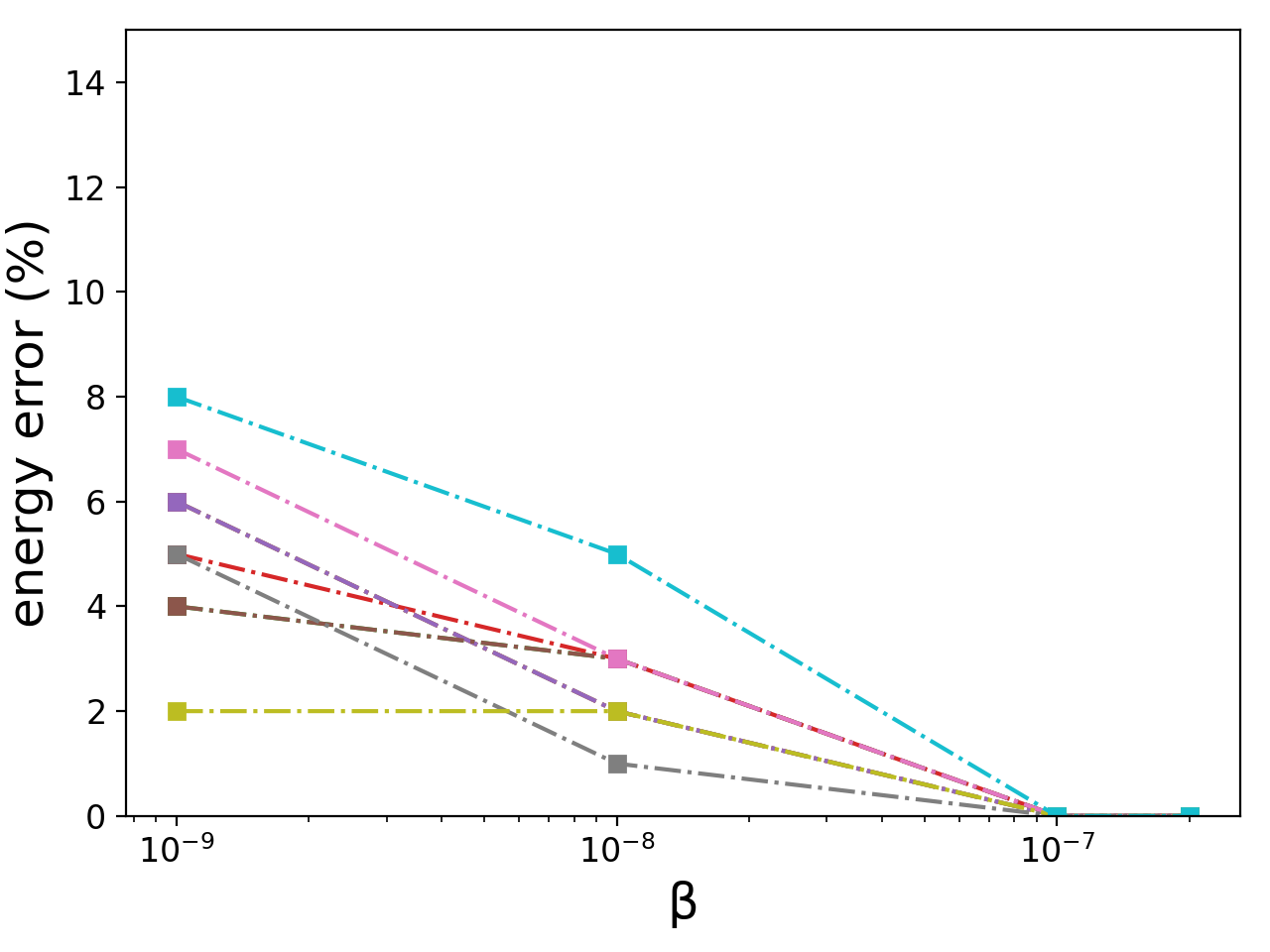} }}
    
    \subfloat[$C_3, L=32$]{{\includegraphics[width=0.47\textwidth]{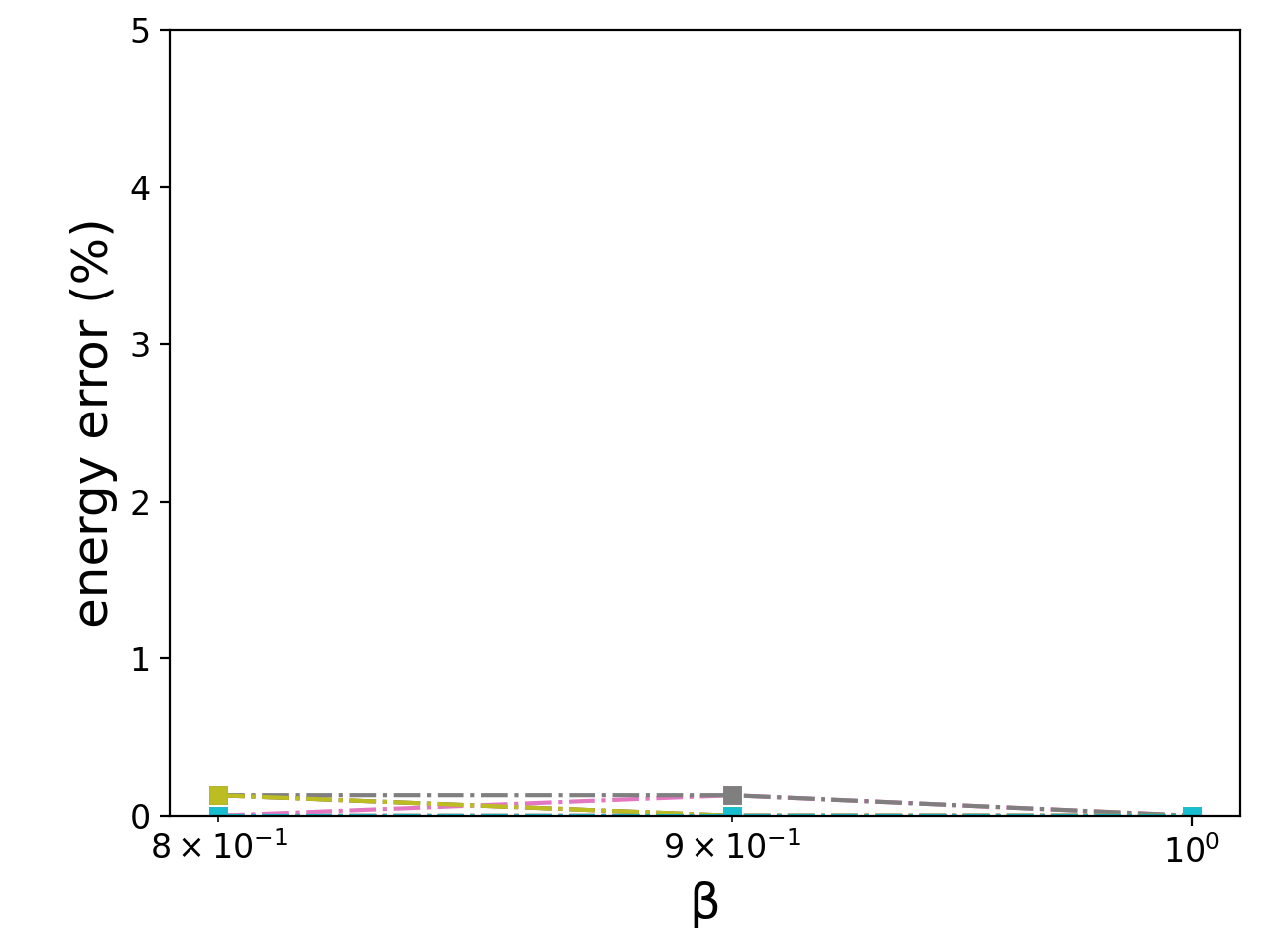} }} \hspace{0.4cm}
    \subfloat[$C_4, L=32$]{{\includegraphics[width=0.47\textwidth]{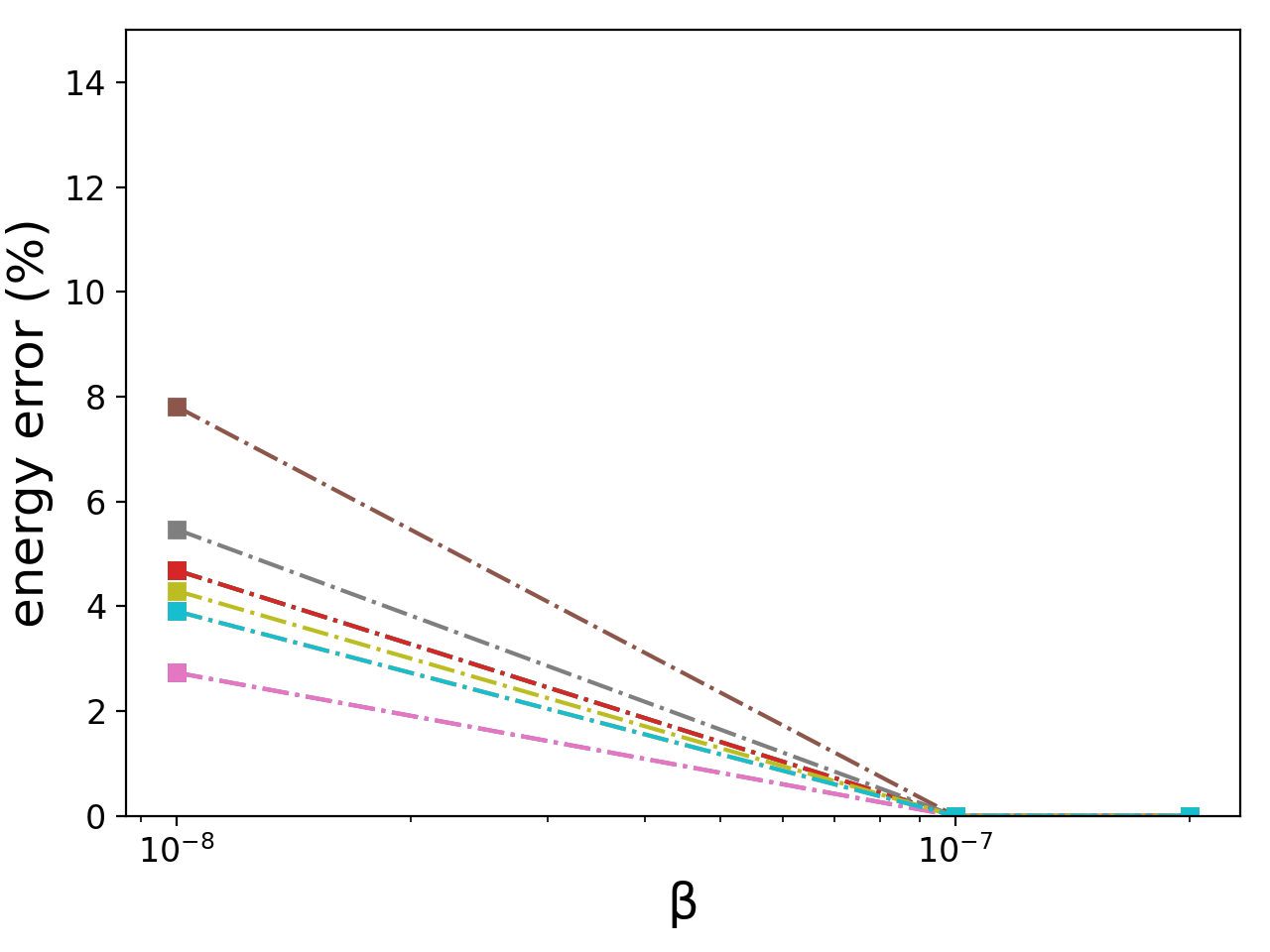} }}

    \subfloat[$C_3, L=48$]{{\includegraphics[width=0.47\textwidth]{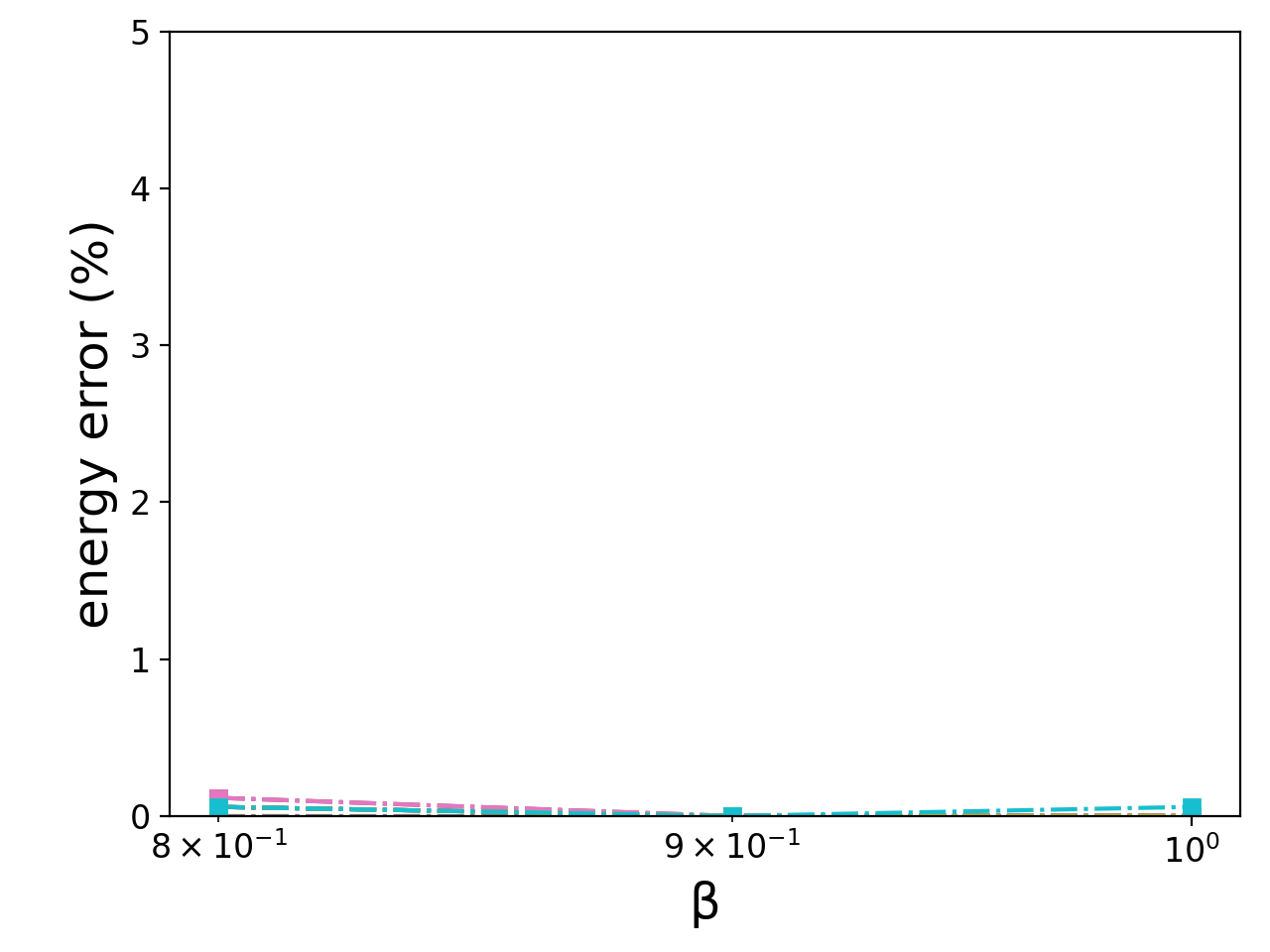} }} \hspace{0.4cm}
    \subfloat[$C_4, L=48$]{{\includegraphics[width=0.47\textwidth]{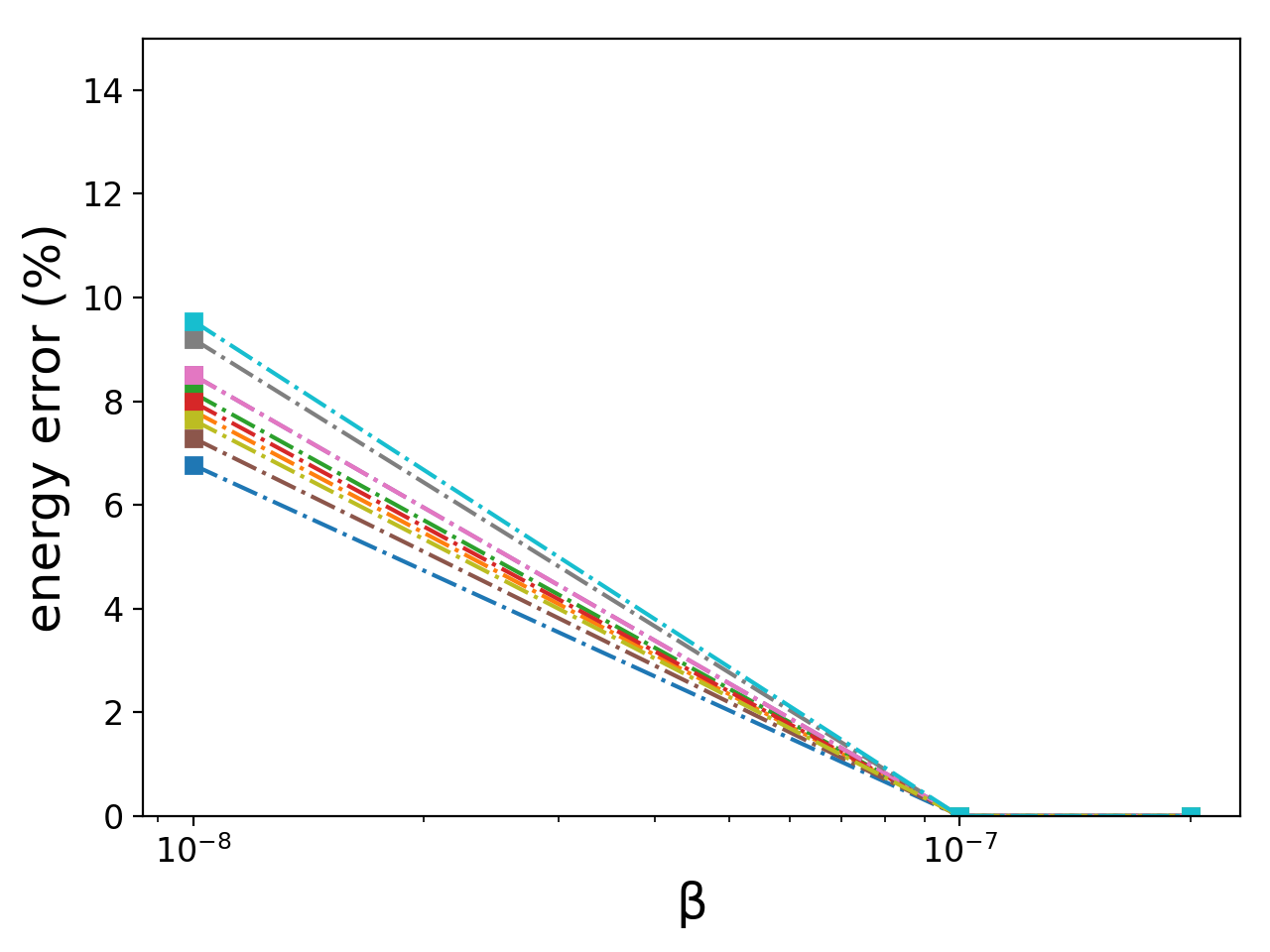} }}
    
    \caption{(color online). $C_3$ (left column) and $C_4$ (right column): solution quality vs. $\beta$, ten instances with $\chi=4$.  Data points belonging to the same instance are connected by dash-dotted lines.}
\end{figure*}

\end{document}